\documentclass[11pt,a4paper]{article}
\usepackage{jheppub}

\usepackage{multirow, graphicx,amssymb,url,mathrsfs,amsmath}
\usepackage{wrapfig,boxedminipage,setspace,subfigure,epsfig}
\usepackage{amsxtra,amstext,latexsym,dsfont,amsfonts}
\usepackage{color,eucal}
\usepackage[dvipsnames]{xcolor}
\usepackage{float}
\usepackage{slashed,comment}
\usepackage{kotex}
\usepackage{tensor}
\usepackage{mathtools}
\usepackage{indentfirst}

\usepackage{array} 
\usepackage[section]{placeins}

\makeatletter

\renewcommand{\maketag@@@}[1]{\hbox{\m@th\normalsize\normalfont#1}}%

\makeatother

\title{The second-order quasi-normal modes for AdS black branes}

\author[a,b]{Wen-Bin Pan,}
\author[b,c]{Zhangping Yu}
\author[b,c]{and Yi Ling,}

\emailAdd{panwb@ihep.ac.cn}
\emailAdd{lingyi@ihep.ac.cn}
\emailAdd{yuzp@ihep.ac.cn}

\affiliation[a]{College of Mathematics and Physics, Beijing University of Chemical Technology, Beijing 100029, China}
\affiliation[b]{Institute of High Energy Physics, Chinese Academy of Sciences, Beijing 100049, China}
\affiliation[c]{School of Physical Sciences, University of Chinese Academy of Sciences, Beijing 100049, China}
	
\abstract{We investigate second-order gravitational perturbations in asymptotically AdS black branes, developing a gauge-invariant framework to compute the amplitude ratio between quadratic and linear quasi-normal modes. Our analysis reveals resonant divergences of this ratio when the summed frequencies of two source modes coincide with the frequency of a third mode. These divergences are shown to manifest as poles in three-point fully retarded correlators of the energy-momentum tensor in the holographically dual quantum field theory, establishing a concrete connection between bulk gravitational nonlinearities and observables in the dual boundary theory. Our findings contribute to the understanding of nonlinearity in quantum many-body systems while deepening the holographic dictionary between spacetime dynamics and quantum correlations.
}

\arxivnumber{}

\begin{document}
 \maketitle
\flushbottom

	
\section{Introduction}\label{sec1}

Linear perturbation theory in general relativity has been pivotal in understanding the dynamics of black hole spacetimes, particularly in describing post-merger ringdown phases of binary black holes \cite{Price:1994pm,Berti:2007fi}, as well as the processes associated with particles plunging into black holes \cite{Zerilli:1970wzz,Davis:1971gg,Berti:2010ce}.
However, the investigation of nonlinear gravitational effects is increasingly recognized as a critical frontier, offering a deeper understanding of the dynamics of black hole mergers and other extreme astrophysical events, where the nonlinearities inherent in Einstein’s equations become pronounced.

The standard paradigm interprets gravitational waves in the ringdown phase as a superposition of quasi-normal modes (QNMs) \cite{Vishveshwara:1970zz,Press:1971wr,Kokkotas:1999bd,Nollert:1999ji,Berti:2009kk,Konoplya:2011qq} $-$ damped oscillations characterized by discrete complex quasi-normal frequencies (QNFs).
While this linear description has long been the cornerstone of ringdown analysis, recent advancements in numerical relativity \cite{Cheung:2022rbm,Mitman:2022qdl} have revealed a more intricate picture, providing compelling evidence for nonlinear mode excitation during the ringdown phase. These quadratic quasi-normal modes(QQNMs) \cite{Gleiser:1995gx,Bruni:1996im,Matarrese:1997ay,Spiers:2023cip,Spiers:2023mor}, arising from the coupling of linear QNMs, exhibit characteristic frequencies
\begin{equation}
	\omega^{(2)}_{n\times m}=\omega^{(1)}_n+\omega^{(1)}_m,
\end{equation}
where the superscripts $(1)$ and $(2)$ denote the linear and quadratic modes, respectively, and the subscripts $n$ and $m$ represent the indices (momenta and overtones) of the coupled linear modes.
Importantly, the quadratic-to-linear amplitude ratio
\begin{equation}
	\mathcal{A}=\frac{A^{(2)}_{n\times m}}{A^{(1)}_n A^{(1)}_m},
\end{equation}
serves as a key observable to quantify the strength of these nonlinear effects 
\cite{Ioka:2007ak,Nakano:2007cj,Lagos:2022otp,Perrone:2023jzq,Bucciotti:2023ets,Khera:2023oyf,Bucciotti:2024zyp,Bourg:2024jme,Bucciotti:2024jrv,Kehagias:2024sgh,Ma:2024qcv,Khera:2024yrk}.
Notably, detecting and characterizing such nonlinear modes offers a powerful probe to test general relativity (GR) and alternative theories of gravity.

However, the exploration of such nonlinear effects in AdS geometries remains sparse, despite their potential to shed light on holographic phenomena such as the nonlinear responses in dual strongly coupled systems.
The AdS/CFT correspondence \cite{Maldacena:1997re} has established a profound connection between the linear QNMs of AdS black holes \cite{Horowitz:1999jd,Cardoso:2001bb,Cardoso:2001vs,Kovtun:2005ev,Miranda:2005qx,Miranda:2008vb,Morgan:2009pn} and the poles of retarded correlators in dual quantum field theories \cite{Birmingham:2001pj,Son:2002sd}. 
Advancing this correspondence into the nonlinear regime represents a natural and compelling direction, as it would deepen our understanding of the role of gravitational nonlinearities in holographic dualities. 
This motivates our investigation of quadratic gravitational QNMs in AdS black branes, where we aim to uncover the holographic significance of these nonlinear modes.
The previous study \cite{Pantelidou:2022ftm} has explored quadratic modes in a model with scalar fields, where nonlinearities arise from cubic polynomial terms such as $\lambda \phi^3$.
The leading nonlinear order gravitational interactions governed by Einstein's equations also manifest as cubic vertices. 
However, the nonlinearity in gravity is intrinsic and structurally dictated by diffeomorphism invariance, leading to new complexity due to gauge redundancy and non-polynomial couplings.

In this work, we compute the amplitude ratio $\mathcal{A}$ for gauge-invariant variables of gravitational fluctuations \cite{Moncrief:1974am,Garat:1999vr,Brizuela:2007zza,Brizuela:2009qd} at the boundary of an AdS$_4$ black brane. 
Our analysis reveals two key findings: (1) $\mathcal{A}$ exhibits resonant divergence under specific frequency conditions of the source modes.
While the current results are explicitly derived in four-dimensional spacetime, we expect that black branes in higher dimensions will exhibit analogous resonances, which come from the planar symmetry of the branes\footnote{A complementary analysis of the QQNMs sourced by tensor modes in AdS$_5$ is presented in Appendix \ref{D}.}.
(2) These resonances correspond to non-trivial poles in three-point energy-momentum fully retarded functions in dual quantum field theory, governing the next-to-linear-order responses on the boundary. 
This correspondence reveals how gravitational nonlinearities in the bulk manifest as measurable quantities in the boundary theory.

The remaining content of this paper is organized as follows. 
In section \ref{sec2}, we discuss about the linear and quadratic order gravitational perturbations. We make a mode decomposition for both modes in section \ref{sec2.1} and derive the equation of motion of the quadratic modes in section \ref{sec2.2}.
In section \ref{sec3}, we study the quadratic-to-linear amplitude in the AdS black brane background.
Section \ref{sec3.1} and \ref{sec3.2} are devoted to our numerical method and numerical results of the amplitude.
Finally in section \ref{sec4}, we summarize this paper and discuss future directions.

\section{The gravitational perturbations over AdS black branes}\label{sec2}

We study gravitational perturbations in the AdS$_4$ black brane background, whose geometry is governed by the metric
\begin{equation}\label{AdS4}
	\mathrm{d}s^2=-r^2f(r)\mathrm{d}t^2+\frac{1}{r^2 f(r)}\mathrm{d}r^2+r^2(\mathrm{d}x^2+\mathrm{d}y^2),
\end{equation}
where $f(r)=1-r_h^3/r^3$, with $r_h$ denoting the radial coordinate of the horizon and the AdS radius normalized to unity.
The Hawking temperature is given by $T=\frac{3 r_h}{4\pi}$. 

Now we consider small gravitational perturbations over the AdS$_4$ black brane background.
Due to the non-linear nature of the Einstein equation, one could explore the dynamics of the perturbations beyond the first linear order.  
As shown in the following, the perturbations at the second-order are sourced by the linear ones. 
One could assume a full metric as 
\begin{equation}\label{ansatz}
	g_{\mu \nu}=\bar{g}_{\mu \nu}+\epsilon\; h_{\mu\nu},
\end{equation}
where $\bar{g}_{\mu \nu}$ denotes the background metric \eqref{AdS4} and $\epsilon \ll 1$ is a bookkeeping parameter that helps to track the order of perturbations.
In the absence of matter, i.e. $T_{\mu \nu}=0$, the metric \eqref{ansatz} should satisfy the vacuum Einstein equation
\begin{equation}\label{Einstein}
	G_{\mu \nu}[\bar{g}+\epsilon \; h]=0,
\end{equation}
where $G_{\mu \nu}[g]=R_{\mu \nu}[g]-\Lambda g_{\mu \nu}$, and the lower indices of the tensors $\bar{g}_{\mu \nu}$ and $h_{\mu\nu}$ are omitted for simplicity.
Note that $\bar{g}+\epsilon \;h$ is a full solution to the Einstein equation rather than a first-order approximate one. 
We expand \eqref{Einstein} with respect to $\epsilon$ up to the second-order:
\begin{equation}
	G_{\mu \nu}[\bar{g}]+\epsilon\; G^{(1)}_{\mu \nu}[h]+\epsilon^2 G^{(2)}_{\mu \nu}[h,h]+\mathcal{O}(\epsilon^3)=0,
\end{equation}
where $G^{(1)}_{\mu \nu}$ and $G^{(2)}_{\mu \nu}$ are respectively linear and bilinear differential operators of $h_{\mu\nu}$. 
The zeroth order term vanishes by definition, while the first-order term is
\begin{equation}\label{linEins}
	G^{(1)}_{\mu \nu}[h]=-\frac{1}{2}\Box h_{\mu\nu} - R_{\mu \alpha \nu \beta} h^{\alpha \beta}+\nabla_\alpha \nabla_{(\mu} \bar{h}^\alpha_{\;\; \nu)}-\Lambda h_{\mu \nu},
\end{equation}
where $\bar{h}^{\mu \nu}$ denotes the trace reversal with respect to the background metric, i.e. $\bar{h}^{\mu \nu}=h^{\mu \nu}-\frac{1}{2}g_{\mu\nu}g^{\alpha \beta}h_{\alpha \beta}$.
Up to the first-order, one only needs to solve the linear equation $G^{(1)}_{\mu \nu}[h]=0$ and take the solution $h^{(1)}_{\mu \nu}$ as the first-order approximation of the full perturbation $h_{\mu \nu}$. Up to the second-order, we have
\begin{equation}\label{quaEins}
	G^{(1)}_{\mu \nu}[h]+\epsilon\; G^{(2)}_{\mu \nu}[h,h]=0.
\end{equation}
Because $\epsilon$ is a small parameter, we expect the solution of \eqref{quaEins} can be written in the form of
\begin{equation}\label{h1+eh2}
	h_{\mu \nu}=h^{(1)}_{\mu \nu}+\epsilon\; h^{(2)}_{\mu \nu},
\end{equation}
i.e. the first-order solution $h^{(1)}_{\mu \nu}$ is corrected by a quantity that is of order $\mathcal{O}(\epsilon)$.
Then after substituting the ansatz \eqref{h1+eh2} into \eqref{quaEins} and noting that $G^{(1)}_{\mu\nu}[h^{(1)}]=0$, we obtain
\begin{equation}\label{quaEinsh}
	G^{(1)}_{\mu \nu}[h^{(2)}]+G^{(2)}_{\mu \nu}[h^{(1)},h^{(1)}]=0.
\end{equation}
Notably, the second-order equation differs from the first-order one only by the inclusion of a source term on the right-hand side, which is quadratic in the first-order solution $h^{(1)}_{\mu\nu}$, while the form of the homogeneous part remains identical except that the variable is changed to be the second-order perturbation $h^{(2)}_{\mu\nu}$.

\subsection{Mode decomposition}\label{sec2.1}

By virtue of the translational symmetry of the AdS black brane, we adopt the ansatz\footnote{The perturbation order $n$ of the wave vectors $k_x, k_y$, and frequency $\omega$ has been omitted for simplicity. One should keep in mind that the wave vectors and frequencies of different orders are generally different.}:
\begin{equation}\label{Fourier}
	h^{(n)}_{\mu \nu}(t,x,y,r)=e^{-i \omega t+ i k_x x + i k_y y} \tilde{h}^{(n)}_{\mu \nu}(\omega,k_x,k_y,r),
\end{equation}
where $n=1,2$ denotes the order of the perturbation.
Given a normalized wave vector $\hat{k}=(k_x/k, k_y/k)$ with $k=\sqrt{k_x^2+k_y^2}$, the tensor $\tilde{h}^{(i)}_{\mu \nu}$ can be decomposed as follows\footnote{A factor $r^2$ is introduced in all terms in the decomposition for later convenience in writing down the master equations.}:
\begin{equation}\label{decom}
	\begin{aligned}
		\tilde{h}^{(n)}_{ab}=&r^2 H^{(n)}_{ab},\\
		\tilde{h}^{(n)}_{ai}=& r^2 H^{(n)}_{a}\hat{k}_{i}+r^2  V^{(n)}_{a}\hat{k}_{\perp i},\\
		\tilde{h}^{(n)}_{ij}=&r^2 H^{(n)}_{L} \delta_{ij}+r^2  H^{(n)}_{T}(\hat{k}_{i}\hat{k}_{j}-\frac{1}{2}\delta_{ij}\hat{k}^2)\\&+r^2  V^{(n)}(\hat{k}_{i}\hat{k}_{\perp j}+\hat{k}_{j}\hat{k}_{\perp i}),		
	\end{aligned}
\end{equation}
where $\hat{k}_{\perp}=(-k_y/k, k_x/k)$ represents a vector perpendicular to $\hat{k}$. In addition, the indices $a, b$ are used to denote coordinates $t$ or $r$, while $i, j$ denote $x$ or $y$. As a result, the variables in the decomposition are classified into two types:
\begin{equation}\label{vecsca}
	\begin{aligned}
		&\text{ Vector:}  &V^{(n)}_{a}, \quad & V^{(n)};\\
		&\text{ Scalar:}  &H^{(n)}_{ab}, \quad & H^{(n)}_{a},  & H^{(n)}_{L}, \quad & H^{(n)}_{T}.
	\end{aligned}
\end{equation}
For $n=1$, these two types of variables are decoupled in the linearized Einstein equation \eqref{linEins} and can be treated separately.
Additionally, for each type of variable, one can construct a master variable that is gauge invariant by combining the variables of that type.
Then Equation \eqref{linEins} can be reduced to two independent ordinary differential equations.

We choose the master variables for $n=1$ to be the Kovtun-Starinets (KS) variables defined in \cite{Kovtun:2005ev}. 
Proceeding further, we also define master variables for $n=2$ to have the same form as for $n=1$, i.e.
\begin{equation}\label{masterV}
	\begin{aligned}
		&\text{ Vector:}  &Y^{(n)} :=&  \mathfrak{q} V^{(n)}_{t}+\mathfrak{w} V^{(n)};\\
		&\text{ Scalar:}  &Z^{(n)} :=& \mathfrak{q}^2 H^{(n)}_{tt}+2 \mathfrak{w} \mathfrak{q} H^{(n)}_{t}+\mathfrak{w}^2 H^{(n)}_{T}+\frac{\left(r^2 f(r)\right)'}{2r}\mathfrak{q}^2\left(H^{(n)}_{L}-\frac{1}{2}H^{(n)}_{T}\right),
	\end{aligned}
\end{equation}
where $\mathfrak{w}=\frac{\omega}{2\pi T}$ and $\mathfrak{q}=\frac{k}{2\pi T}$ are the dimensionless frequency and the magnitude of the wave vector, respectively.
The second-order ODEs for $Y^{(1)}$ and $Z^{(1)}$ are respectively given by
	\begin{small}
		\begin{equation}\label{Master1stY}
			Y''(r) + \left(\frac{4}{r}+\frac{\mathfrak{w}^2 f'(r)}{f(r) (\mathfrak{w}^2 - \mathfrak{q}^2 f(r))}\right) Y'(r) + \left(\frac{9 r_h^2(\mathfrak{w}^2 - \mathfrak{q}^2 f(r))}{4r^4 f(r)^2}\right) Y(r)=0,
		\end{equation}
		\begin{equation}\label{Master1stZ}
			\begin{aligned}
				Z''(r) +\left(\frac{f'(r)}{ f(r)}+\frac{10}{r}+ \frac{24(\mathfrak{q}^2-\mathfrak{w}^2)}{r(4 \mathfrak{w}^2-3 \mathfrak{q}^2 - \mathfrak{q}^2 f(r))}\right)Z'(r) + \left(\frac{\mathfrak{q}^2 f'(r)^2}{f(r) (4 \mathfrak{w}^2 - 3 \mathfrak{q}^2 - \mathfrak{q}^2 f(r))}\right.\\
				\left.+\frac{9 r_h^2(\mathfrak{w}^2-\mathfrak{q}^2 f(r))}{4 r^4 f(r)^2 }\right) Z(r)=0.
			\end{aligned}		
		\end{equation}
	\end{small}
Imposing the ingoing boundary condition at the horizon and the Dirichlet boundary condition at infinity, one may solve the above master equations to figure out the quasi-normal modes of the black brane. These modes are characterized by a discrete set of quasi-normal frequencies $\omega_n$.

To recover the metric components from the master variables, one must fix a gauge. 
Under an infinitesimal diffeomorphism $x^\mu \to x^\mu+\epsilon \xi^\mu$, first-order metric fluctuations transform as 
\begin{equation}
	\begin{aligned}
		h_{\mu \nu}^{(1)} \rightarrow h_{\mu \nu}^{(1)}+\mathcal{L}_{\xi} \bar{g}_{\mu \nu}=h_{\mu \nu}^{(1)}+2 \nabla_{(\mu} \xi_{\nu)},
	\end{aligned}
\end{equation} 
$Y^{(1)}$ and $Z^{(1)}$ are both invariant under such transformations. We choose the Regge-Wheeler gauge \cite{Regge:1957td,Zerilli:1970se} for the first-order perturbation $h^{(1)}_{\mu\nu}$, which imposes conditions
\begin{equation}
	V^{(1)}=H^{(1)}_{a}=H^{(1)}_{T}=0.
\end{equation}
Then the remaining variables of vector and scalar type can be reconstructed by the gauge invariants $Y^{(1)}(r)$ and $Z^{(1)}(r)$ respectively. 
Specifically, the reconstruction for the variables in the vector sector is
\begin{equation}\label{reconst}
	\begin{aligned}
		V^{(1)}_{t}=&\frac{1}{\mathfrak{q}}Y^{(1)}(r)\\
		V^{(1)}_{r}=&\frac{i \mathfrak{w}}{2\pi T \mathfrak{q}\left(\mathfrak{w}^2-\mathfrak{q}^2f(r)\right)} \frac{\mathrm{d} Y^{(1)}(r)}{\mathrm{d} r}.
	\end{aligned}
\end{equation}
The detailed derivation of this reconstruction is presented in Appendix \ref{A}. 

\subsection{The quadratic quasi-normal modes}\label{sec2.2}
Next, we derive the equation of motion for the quadratic quasi-normal modes from equation \eqref{quaEinsh} with selected two linear-order quasi-normal modes as the source.
Equation \eqref{quaEinsh} can be explicitly written as
\begin{equation}\label{quadeq}
	\delta R_{\mu\nu}[h^{(2)}]-\Lambda h^{(2)}_{\mu\nu}+\delta^2 R_{\mu\nu}[h_1^{(1)},h_2^{(1)}]=0
\end{equation}
where the source term is \cite{Spiers:2023mor}
\begin{equation}
	\begin{aligned}
		\delta^2 R_{\mu\nu}[h_1^{(1)},h_2^{(1)}]=&\frac{1}{2}\left(A_{\alpha \beta}[h_1^{(1)},h_2^{(1)}]+B_{\alpha \beta}[h_1^{(1)},h_2^{(1)}]+C_{\alpha \beta}[h_1^{(1)},h_2^{(1)}] \right),
	\end{aligned}
\end{equation}
and
\begin{equation}
	\begin{aligned}
		A_{\alpha \beta}[h] & :=\frac{1}{2} h^{\mu \nu}{ }_{; \alpha} h_{\mu \nu ; \beta}+h^{\mu}{ }_{\beta}{ }^{;\nu}\left(h_{\mu \alpha ; \nu}-h_{\nu \alpha ; \mu}\right), \\
		B_{\alpha \beta}[h] & :=-h^{\mu \nu}\left(2 h_{\mu(\alpha ; \beta) \nu}-h_{\alpha \beta ; \mu \nu}-h_{\mu \nu ; \alpha \beta}\right), \\
		C_{\alpha \beta}[h] & :=-\bar{h}^{\mu \nu}{ }_{; \nu}\left(2 h_{\mu(\alpha ; \beta)}-h_{\alpha \beta ; \mu}\right).
	\end{aligned}
\end{equation}
Note that the source is quadratic in $h^{(1)}$. 

Plugging the ansatz for the linear-order modes in \eqref{Fourier} into \eqref{quadeq} and supposing that the frequency and wave vector of the two source modes $h_1^{(1)}$ and $h_2^{(1)}$ are $(\omega_1, \vec{k}_1)$ and $(\omega_2, \vec{k}_2)$, respectively, then the frequency $\omega$ and the wave vector $\vec{k}$ for the quadratic-order mode $h^{(2)}$ have to be
\begin{equation}\label{sum}
	(\omega=\omega_1+\omega_2,\; \vec{k}=\vec{k}_1+\vec{k}_2),
\end{equation}
where $\omega_1$ and $\omega_2$ are quasi-normal frequencies that depend on continuous parameters $\vec{k}_1$ and $\vec{k}_2$, respectively.
We remark that if the quadratic-order mode is sourced by two distinct linear-order modes, then there are two possible contributions in the source \footnote{One contribution comes from $\tilde{h}_1^{(1)}(\omega_1, \vec{k}_1, r)$ and $\tilde{h}_2^{(1)}(\omega_2, \vec{k}_2, r)$, the other comes from $\tilde{h}_1^{(1)}(\omega_2, \vec{k}_2, r)$ and $\tilde{h}_2^{(1)}(\omega_1, \vec{k}_1, r)$.}. However, if two source modes are identical, then there is only one contribution.
Without loss of generality, we choose the wave vector of the quadratic-order mode $\vec{k}$ to be along the $x$ direction, i.e. $\vec{k}=(k_x, 0)$.
Therefore, we could write $\vec{k}_1=(k_{1x}, k_y)$ and $\vec{k}_2=(k_{2x}, -k_y)$.

The vector and scalar types of variables in $h^{(2)}$ are also decoupled in equation \eqref{quadeq} because its dynamical part is identical to the linear order (i.e. $G^{(1)}_{\mu\nu}$). 
In fact, with the choice of $\vec{k}$ described above, equations that consist of a dynamical part $G^{(1)}_{ty}$, $G^{(1)}_{ry}$ or $G^{(1)}_{xy}$ determine vector modes while the others determine scalar modes.
Therefore, equation \eqref{quadeq} can be reduced to two inhomogeneous ODEs for each of the master variables $Y^{(2)}$ and $Z^{(2)}$, with the same homogeneous parts as in \eqref{Master1stY} and \eqref{Master1stZ}.
The master equation has the following form,
\begin{equation}\label{Master2nd}	
	(\psi^{(2)})''(r)+p(r)(\psi^{(2)})'(r)+q(r)\psi^{(2)}(r)=S[\tilde{h}_1^{(1)}, \tilde{h}_2^{(1)}],
\end{equation}
where $\psi^{(2)}$ denotes $Y^{(2)}$ or $Z^{(2)}$, and $S$ represents the source term.

Then, we also represent $\tilde{h}_1^{(1)}$ and $\tilde{h}_2^{(1)}$ using the master variables through the reconstruction of metric perturbations\footnote{It should be emphasized that one could fix the RW gauge for both of the source modes. The details are shown in Appendix \ref{A}.}. The resulting form of the source is
\begin{equation}\label{Source2nd}
	\begin{aligned}
		S[\psi_1, \psi_2]=&F_1 \psi_1'(r) \psi_2'(r)+F_2 \psi_1'(r) \psi_2(r)+F_3 \psi_1(r) \psi_2'(r)+F_4 \psi_1(r) \psi_2(r),
	\end{aligned}
\end{equation}
where $\psi_1$ and $\psi_2$ are two first-order master variables, each of which represents $Y^{(1)}$ or $Z^{(1)}$.
The coefficients $F_j$ $(j=1,2,3,4)$ are functions of $(\omega_1, \vec{k}_1, r)$ and $(\omega_2, \vec{k}_2, r)$, the explicit form of which are presented in Appendix \ref{B}.
Note that we have eliminated the second and higher order derivatives of $\psi_1$ and $\psi_2$ in the source by using first-order master equations \eqref{Master1stY} and \eqref{Master1stZ}.
Since $1$ specific quadratic-order mode is generated by $2$ linear-order modes in the source and each mode has $2$ possible types, we have $6$ independent processes in total:
\begin{equation}\label{process}
	\begin{aligned}   
		Y^{(1)}_1\times Y^{(1)}_2 \to Y^{(2)}, \quad
		Y^{(1)}_1\times Y^{(1)}_2 \to Z^{(2)}, \\
		Z^{(1)}_1\times Z^{(1)}_2 \to Y^{(2)}, \quad
		Z^{(1)}_1\times Z^{(1)}_2 \to Z^{(2)},\\
		Y^{(1)}_1\times Z^{(1)}_2 \to Y^{(2)}, \quad
		Y^{(1)}_1\times Z^{(1)}_2 \to Z^{(2)}.
	\end{aligned}
\end{equation}

\section{The quadratic-to-linear amplitudes}\label{sec3}

After all the manipulations above, we are ready to numerically solve the inhomogeneous equation of \eqref{Master2nd} with the source \eqref{Source2nd}, imposing the ingoing boundary condition at the horizon and the Dirichlet condition at the AdS boundary. 

A general solution satisfying the ingoing boundary condition at the horizon for either a first- or second-order master variable admits the following asymptotic expansion near the AdS boundary:
\begin{equation}\label{nearbdry}
	\begin{aligned}
		\psi^{(n)}(r,\omega,\vec{k})=r^{-(d-\Delta)} \left( A^{(n)}(\omega,\vec{k})+\mathcal{O}(r^{-1})\right) + r^{-\Delta} \left(B^{(n)}(\omega,\vec{k})+\mathcal{O}(r^{-1})\right),
	\end{aligned}
\end{equation}
where the scaling dimension is $\Delta=d=3$.
The Dirichlet boundary condition requires $A^{(n)}=0$ for solutions of QNMs and QQNMs.
Therefore, for fixed source QNFs $(\omega_1,\omega_2)$ and wave vectors $(\vec{k}_1,\vec{k}_2)$, we define the quadratic-to-linear amplitude ratio at the boundary as
\begin{equation}\label{QQNMA}
	\begin{aligned}
		\mathcal{A}&=\lim_{r \to \infty}  \frac{ \psi^{(2)}(r, \omega_1+\omega_2,\vec{k}_1+\vec{k}_2)}{r^3\psi^{(1)}(r, \omega_1,\vec{k}_1)\psi^{(1)}(r, \omega_2,\vec{k}_2)}\\
		&=\frac{B^{(2)}(\omega_1+\omega_2,\vec{k}_1+\vec{k}_2)}{B^{(1)}(\omega_1,\vec{k}_1)B^{(1)}(\omega_2,\vec{k}_2)}.
	\end{aligned}   
\end{equation}
This amplitude ratio is uniquely determined by the master equation \eqref{Master2nd}, in contrast to the amplitudes of linear QNMs, which depend on the initial conditions of the perturbation.

We apply an adapted version of the Horowitz-Hubeny(H-H) method \cite{Horowitz:1999jd} and the pseudo-spectral method to compute the ratio of amplitudes at the AdS boundary.
The results obtained from the two methods are in excellent agreement. 
The H-H method is presented in the following while the pseudo-spectral method is concisely demonstrated in Appendix \ref{C}.

\subsection{Numerical computation}\label{sec3.1}

\paragraph{The linear order modes}
Firstly, we solve for the quasi-normal frequencies and near-horizon series solutions of linear modes using the H-H method.
It is convenient to perform a coordinate transformation $u=\frac{r_h}{r}$, which brings the AdS boundary to $u=0$ and horizon to $u=1$.
We assume that the near-horizon expansion of a master variable $\psi^{(1)}$ has the form
\begin{equation}\label{linmode}
	\begin{aligned}
		\psi^{(1)}(u)=(u-1)^{\lambda}\sum_{j=0}^{+\infty} a^{(1)}_j (u-1)^j.
	\end{aligned}
\end{equation}
We also perform a near horizon expansion for the coefficients $p(u)$ and $q(u)$ in the master equation \eqref{Master1stY},
\begin{equation}\label{pq}
	\begin{aligned}
		p(u)=&(u-1)^{-1}\sum_{j=0}^{+\infty} p_j (u-1)^j\\
		q(u)=&(u-1)^{-2}\sum_{j=0}^{+\infty} q_j (u-1)^j,
	\end{aligned}
\end{equation}
where the leading exponents are $-1$ and $-2$ for $p(u)$ and $q(u)$ respectively.
Plugging \eqref{linmode} and \eqref{pq} into the master equation \eqref{Master1stY}, one finds that it becomes a series expansion of the form
\begin{equation}\label{expeom}
	\begin{aligned}
		\sum^{+\infty}_{j=0} \mathcal{W}^{(1)}_j (u-1)^{\lambda+j-2}= 0.
	\end{aligned}
\end{equation}
The leading order equation reads as $\mathcal{W}^{(1)}_0=(\lambda(\lambda-1)+p_0 \lambda +q_0)a^{(1)}_0=0$.
Combining this with the ingoing boundary condition at the horizon, we obtain $\lambda=-\frac{i \mathfrak{w}}{2}$.
One can then find $a_j$ by solving \eqref{expeom} order by order.
The coefficients with $j\ge 1$ are functions of $(\mathfrak{w},\mathfrak{q})$ and are all proportional to $a_0$, which is the amplitude of the linear order mode.
The boundary condition at $u=0$ imposes further that 
\begin{equation}
	\sum^{+\infty}_{j=0} a^{(1)}_j (-1)^j=0. 
\end{equation}
We truncate the sum at a finite order $N$ and obtain the quasi-normal frequencies numerically.

The quasi-normal frequencies form a discrete complex spectrum for the black brane and can be labeled with an overtone number $n$ along with the sign of their real parts. 
For the vector modes in AdS, there is a hydrodynamic mode whose frequency lies on the negative imaginary axis and approaches $0$ as its wave vector approaches $0$, which is denoted as $n=0$ in the following.
Other non-hydro modes, whose frequencies have positive real parts, are denoted as $n=1,2,3,...$, in the order of their imaginary parts starting from the least damped mode.
Their mirror modes\footnote{The mirror mode of a QNM with frequency $\omega$ has frequency $-\omega^*$.} are labeled as $n=1c,2c,3c,...$, respectively.
Similarly, for the scalar modes, there is a pair of hydrodynamic modes that are labeled by $0$ and $0c$.
Other non-hydro modes are denoted as $n=1,2,3,...$ and $n=1c,2c,3c,...$.

\paragraph{The quadratic order modes}
Next, we solve for the quadratic modes. 
The master equation has the form of 
\begin{equation}\label{Master2ndu}
	\begin{aligned}
		&(\psi^{(2)})''+p(u)(\psi^{(2)})'+q(u)\psi^{(2)}=S[\psi^{(1)}_1, \psi^{(1)}_2],
	\end{aligned}	
\end{equation}
where a prime represents the derivative of $u$.
We also assume a series ansatz for $\psi^{(2)}$ 
\begin{equation}\label{psi2}
	\psi^{(2)}(u)=(u-1)^{\lambda}\sum_{i=0}^{N} a^{(2)}_i (u-1)^i.
\end{equation}
This series ansatz and the series solutions of two linear vector modes are substituted into \eqref{Master2ndu} which is then further expanded up to the truncation order $N$.
It is found that the leading exponent of the series of the source on the right hand side of \eqref{Master2ndu} is $\lambda-1$, which holds for both processes, while that of the left hand side is $\lambda-2$.
Thus, the presence of the source does not change the leading behavior of the quadratic modes $\psi$.
Therefore, the expansion of \eqref{Master2ndu} has the form of
\begin{equation}\label{expeom2nd}
	\begin{aligned}
		(u-1)^{\lambda-2} \left(\mathcal{W}^{(2)}_0+\sum^{N}_{j=1}(\mathcal{W}^{(2)}_j-\mathcal{S}_j)(u-1)^j\right)= 0,
	\end{aligned}
\end{equation}
where $\mathcal{S}_j$ is the contribution from the source.
One may solve it order by order, similar to the linear order. 
By imposing the ingoing boundary condition at the horizon and taking into consideration that $\omega=\omega_1+\omega_2$, the leading equation $\mathcal{W}^{(2)}_0$ gives us
\begin{equation}
	\lambda=\lambda_1+\lambda_2=\frac{-i (\mathfrak{w}_1+\mathfrak{w}_2)}{2}.
\end{equation}
Then $a^{(2)}_j$ can be solved recursively from the $j$-th equation:
\begin{equation}
	\mathcal{W}^{(2)}_j-\mathcal{S}_j=\left(\sum^{j}_{i=0} c_{ji} a^{(2)}_i\right)-b_j s_{j-2} a^{(1)}_0 a^{(1)}_0=0,
\end{equation}
where $j=1,2,3,...,N$, $c_{ji}$ and $b_j$ are constants depending on $\mathfrak{w}$ and $\mathfrak{q}$, and $s_{j-2}$ is the $(j-2)$-th coefficient in the series expansion of the source, which is also a function of $\mathfrak{w}$ and $\mathfrak{q}$.
The result for $a^{(2)}_j$ is in terms of the quadratic-order amplitude at horizon $a^{(2)}_0(\mathfrak{w},\mathfrak{q})$ and the product of two linear-order amplitudes $a^{(1)}_0(\mathfrak{w}_1,\mathfrak{q}_1) a^{(1)}_0(\mathfrak{w}_2,\mathfrak{q}_2)$. 
The vanishing condition at $u=0$ requires that
\begin{equation}
	\sum^{N}_{j=0} a^{(2)}_j (-1)^j=0, 
\end{equation}
which determines a ratio between the amplitudes on the horizon: 
\begin{equation}	\mathcal{A}_N(r_h)=\frac{a^{(2)}_0(\mathfrak{w},\mathfrak{q})}{a^{(1)}_0(\mathfrak{w}_1,\mathfrak{q}_1)  a^{(1)}_0(\mathfrak{w}_2,\mathfrak{q}_2)}
\end{equation}
for each truncation order $N$.

To calculate the amplitude ratio at the AdS asymptotic boundary, we simply substitute $a^{(2)}_0$ in terms of $\mathcal{A}(r_h)$ and $a^{(1)}_0$ into the series expression for the quadratic-order mode in equation \eqref{psi2}, and compute the ratio according to the definition of the boundary amplitude given in equation \eqref{QQNMA}.

\begin{figure}[hbpt]
	\centering
	\includegraphics[scale=0.4]{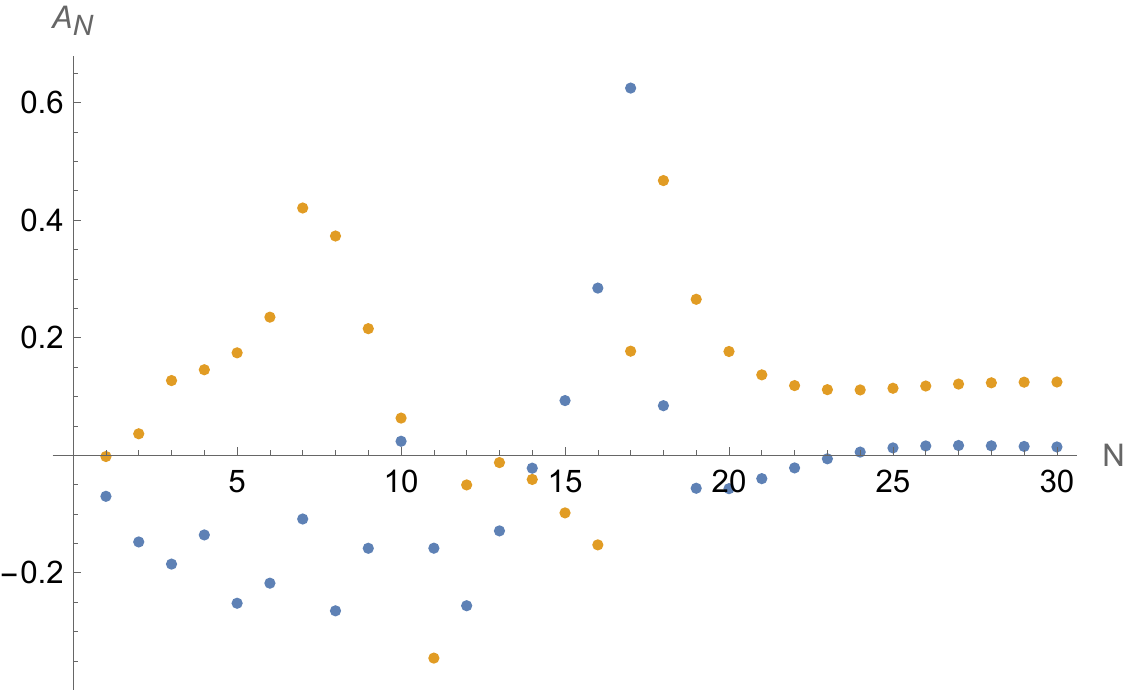}
	\caption{The solution of the quadratic-to-linear ratio of amplitudes $\mathcal{A}_N$ with truncation orders $N=1,2,...,30$, in the process $Y^{(1)}_1\times Y^{(1)}_2 \to Y^{(2)}$ and two source modes are chosen to be $n=1$ and $n=2$. The wave vectors of two source modes are $\Vec{k}_1=(\frac{\sqrt{3}}{2}, \frac{1}{2})$ and $\Vec{k}_2=(\frac{\sqrt{3}}{2}, -\frac{1}{2})$, respectively. The blue and yellow dots represent the real and imaginary parts of $\mathcal{A}_N$, respectively. As $N$ increases, $\mathcal{A}_N$ approaches a fixed complex number. The horizon radius $r_h$ is set to be $1$.
	}\label{converge}
\end{figure}

As $N$ increases, the convergence of the series $\mathcal{A}_N$ can be justified numerically for all the cases under consideration.
The quadratic-to-linear ratio for a typical pattern of $Y^{(1)}_1\times Y^{(1)}_2 \to Y^{(2)}$ is illustrated in Figure (\ref{converge}). 
To obtain a desirable accuracy, we typically need to choose at least $N=30$ for $\left | \vec{k}_1 \right |<4$ and $\left | \vec{k}_2 \right |<4$.
For larger magnitudes of the wave numbers, $N$ must be larger.
We fix $N=30$ in our numerical computations.

\subsection{Numerical results of the amplitude}\label{sec3.2}

\begin{figure}[hbpt]
	\centering
	\subfigure[]{
		\begin{minipage}[b]{.45\linewidth}
			\centering
			\includegraphics[scale=0.35]{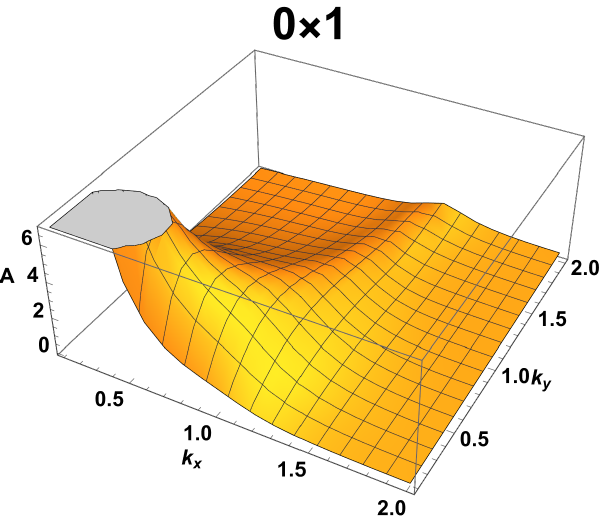}
		\end{minipage}
	}
	\subfigure[]{
		\begin{minipage}[b]{.45\linewidth}
			\centering
			\includegraphics[scale=0.35]{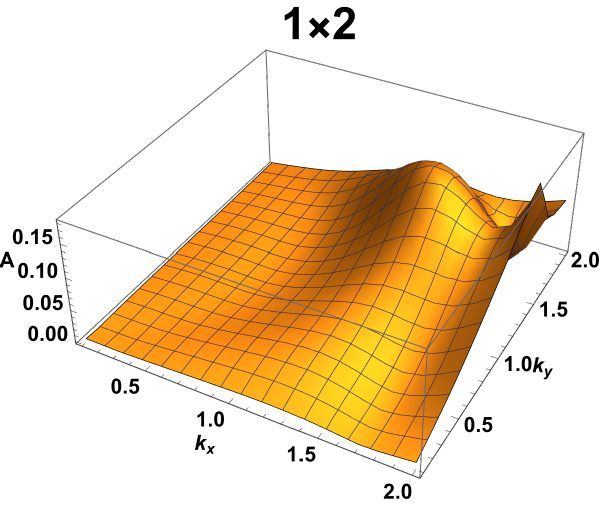}
		\end{minipage}
	}\\
	\subfigure[]{
		\begin{minipage}[b]{.45\linewidth}
			\centering
			\includegraphics[scale=0.35]{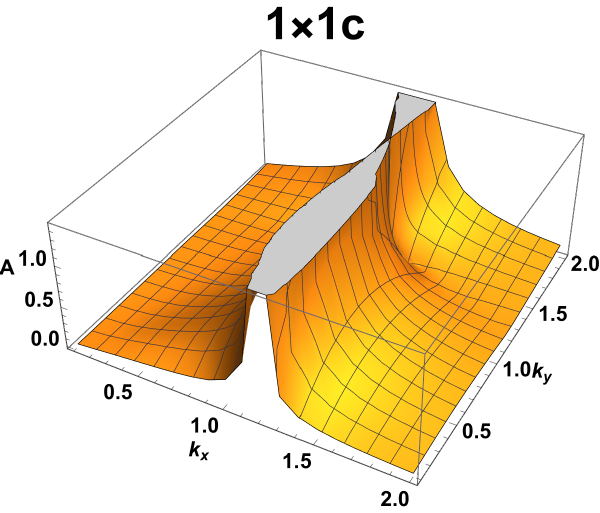}
		\end{minipage}
	}
	\subfigure[]{
		\begin{minipage}[b]{.45\linewidth}
			\centering
			\includegraphics[scale=0.35]{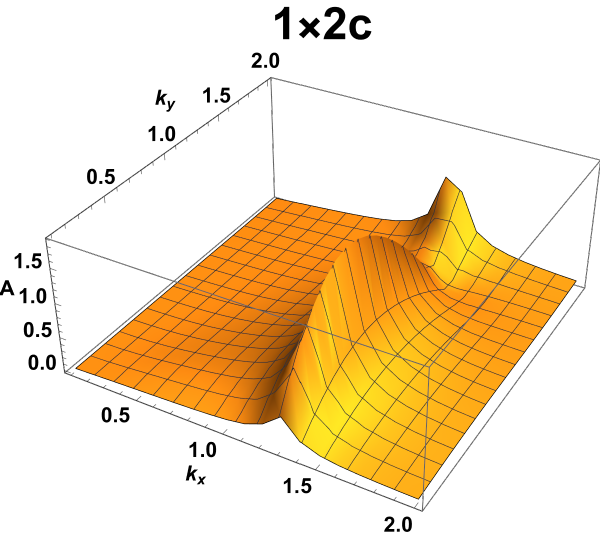}
		\end{minipage}
	}
	\caption{The absolute value of the ratio $\mathcal{A}$ in the case of $Y^{(1)}_1\times Y^{(1)}_2 \to Y^{(2)}$. Each subfigure is labeled by the overtone numbers of the source modes in the format of $n_1 \times n_2$.
    In (a), $\mathcal{A}$ diverge at the origin.
    In (c), there is a divergent curve located approximately at $k_x\approx1.1809$. 
    On this curve, the sum of the QNFs of a pair of mirror modes ($n=1, m=1c$) equals the QNF of the hydrodynamic mode ($l=0$).}
	\label{VVV}
\end{figure}

\begin{figure}[hbpt]
	\centering
	\subfigure[]{
		\begin{minipage}[b]{.45\linewidth}
			\centering
			\includegraphics[scale=0.35]{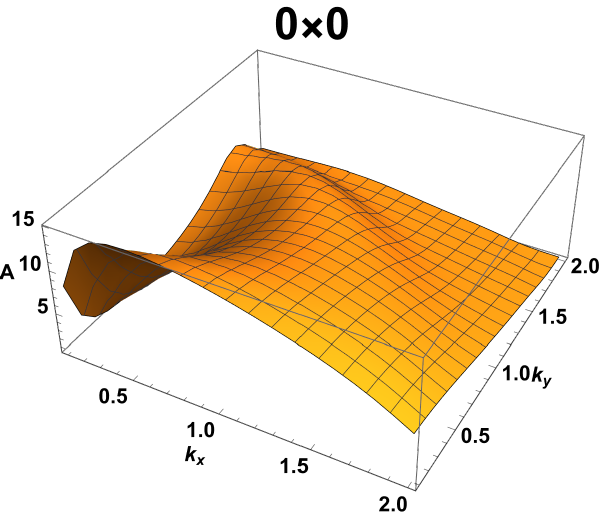}
		\end{minipage}
	}
	\subfigure[]{
		\begin{minipage}[b]{.45\linewidth}
			\centering
			\includegraphics[scale=0.35]{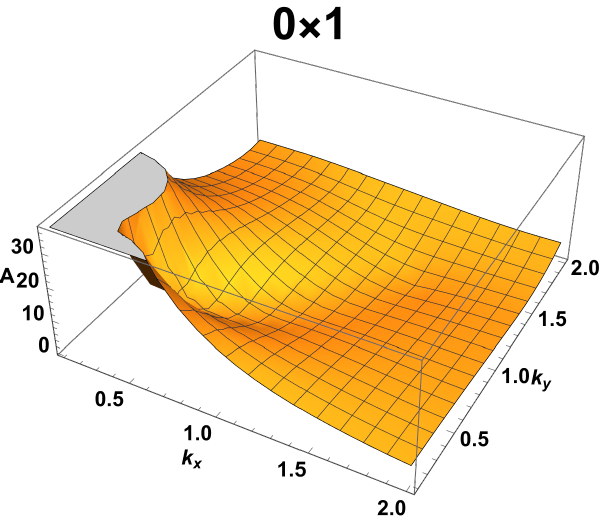}
		\end{minipage}
	}\\
	\subfigure[]{
		\begin{minipage}[b]{.45\linewidth}
			\centering
			\includegraphics[scale=0.35]{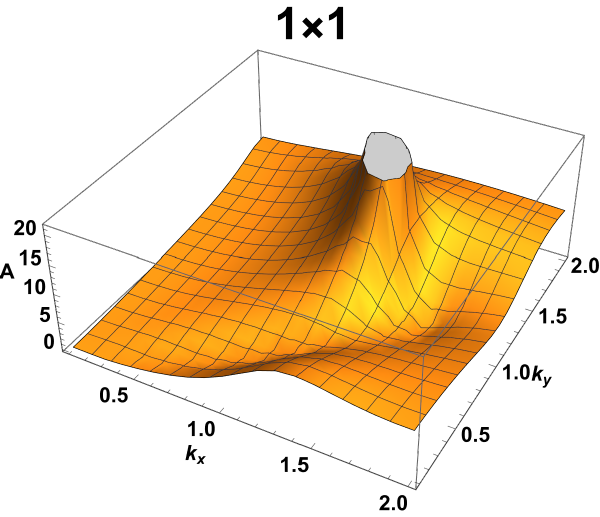}
		\end{minipage}
	}
	\subfigure[]{
		\begin{minipage}[b]{.45\linewidth}
			\centering
			\includegraphics[scale=0.35]{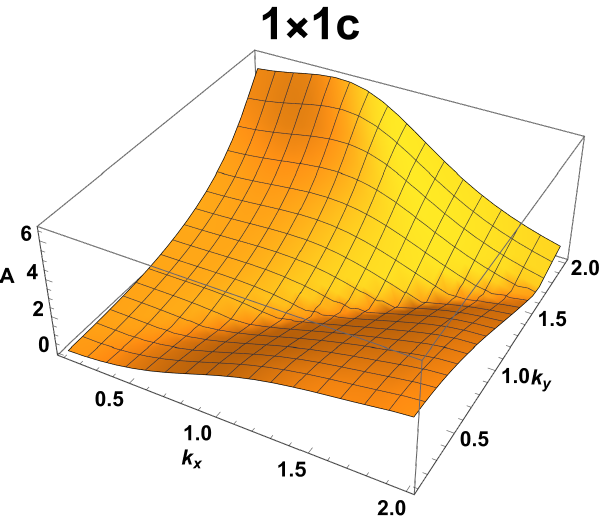}
		\end{minipage}
	}
	\caption{The absolute value of the ratio $\mathcal{A}$ in the case of $Y^{(1)}_1\times Y^{(1)}_2 \to Z^{(2)}$. Each subfigure is labeled by the overtone numbers of the source modes in the format of $n_1 \times n_2$.
    In (b), $\mathcal{A}$ diverge at the origin.
    In (c), there is an isolated resonant point $(k_{x}, k_y)\approx(1.2313, 1.3751)$.
    The corresponding frequencies are $\omega^{(1)}_{n=1}=\omega^{(1)}_{m=1}=\frac{1}{2}\omega^{(1)}_{l=2}\approx 2.0257-2.3890i$.} 
	\label{VVS}
\end{figure}

\begin{figure}[hbpt]
	\centering
	\subfigure[]{
		\begin{minipage}[b]{.45\linewidth}
			\centering
			\includegraphics[scale=0.35]{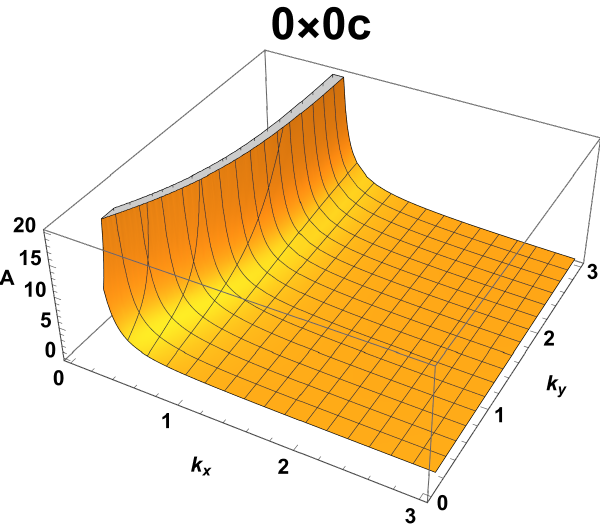}
		\end{minipage}
	}
	\subfigure[]{
		\begin{minipage}[b]{.45\linewidth}
			\centering
			\includegraphics[scale=0.35]{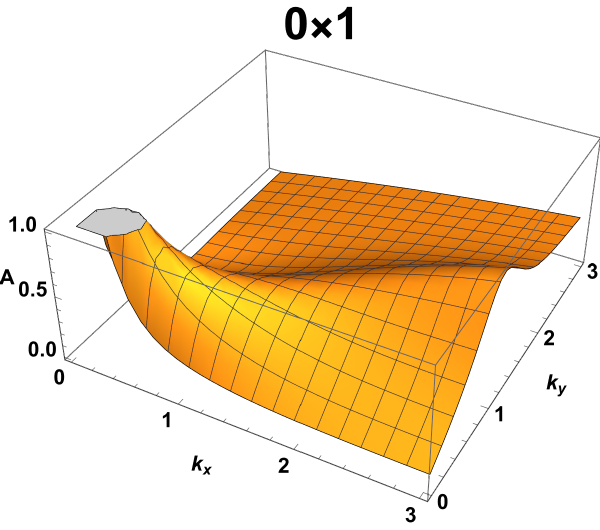}
		\end{minipage}
	}\\
	\subfigure[]{
		\begin{minipage}[b]{.45\linewidth}
			\centering
			\includegraphics[scale=0.35]{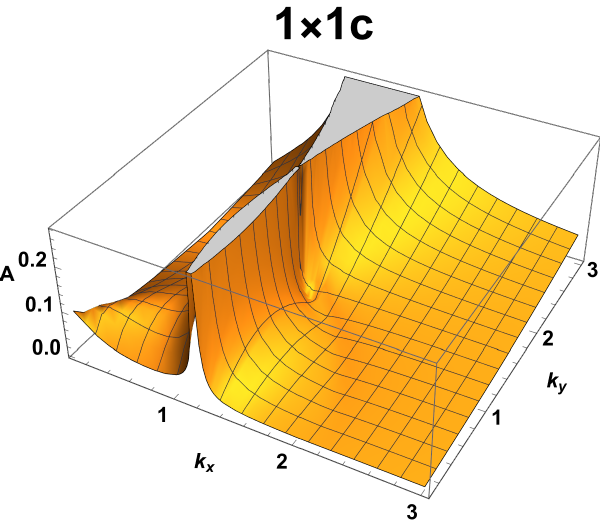}
		\end{minipage}
	}
	\subfigure[]{
		\begin{minipage}[b]{.45\linewidth}
			\centering
			\includegraphics[scale=0.35]{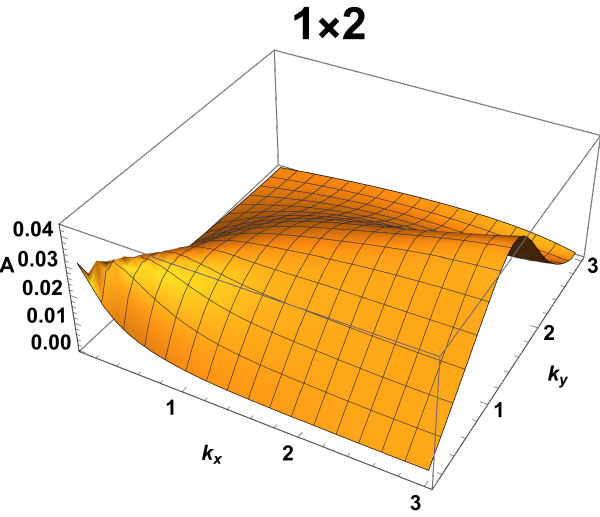}
		\end{minipage}
	}
	\caption{The absolute value of the ratio $\mathcal{A}$ in the case of $Z^{(1)}_1\times Z^{(1)}_2 \to Y^{(2)}$. Each subfigure is labeled by the overtone numbers of the source modes in the format of $n_1 \times n_2$.
    In (b), $\mathcal{A}$ diverge at the origin.
    In (a) and (c), divergent curves are observed.} 
	\label{SSV}
\end{figure}

\begin{figure}[hbpt]
	\centering
	\subfigure[]{
		\begin{minipage}[b]{.45\linewidth}
			\centering
			\includegraphics[scale=0.35]{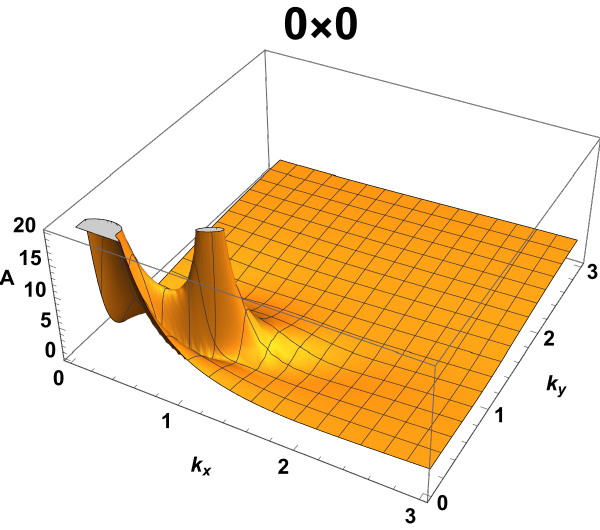}
		\end{minipage}
	}
	\subfigure[]{
		\begin{minipage}[b]{.45\linewidth}
			\centering
			\includegraphics[scale=0.35]{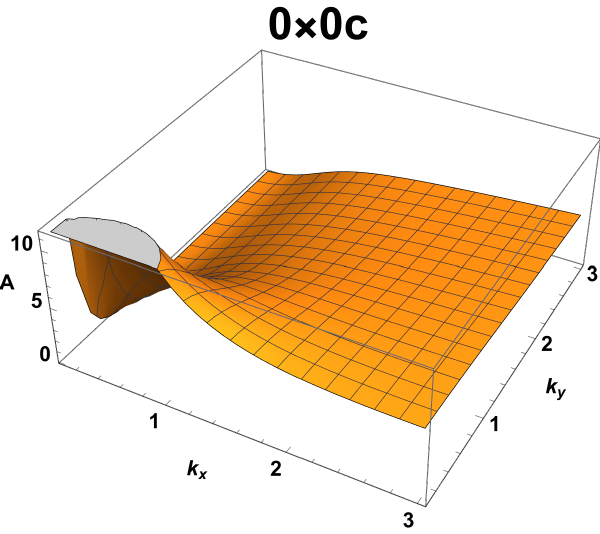}
		\end{minipage}
	}\\
	\subfigure[]{
		\begin{minipage}[b]{.45\linewidth}
			\centering
			\includegraphics[scale=0.35]{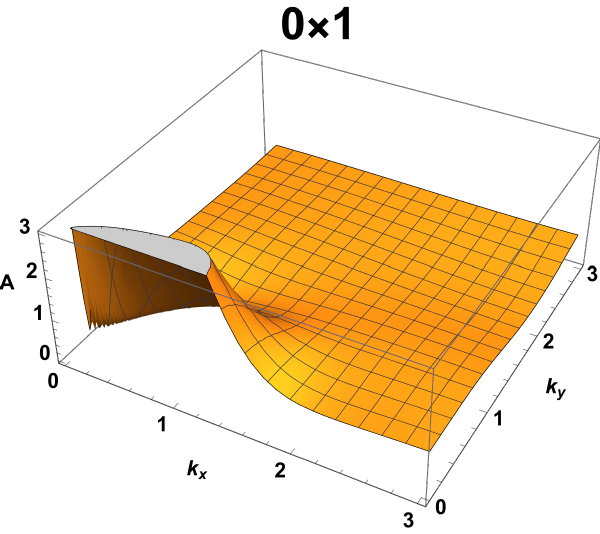}
		\end{minipage}
	}
	\subfigure[]{
		\begin{minipage}[b]{.45\linewidth}
			\centering
			\includegraphics[scale=0.35]{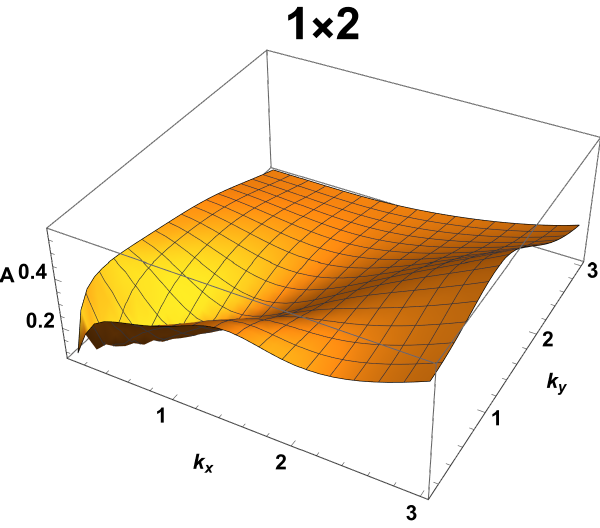}
		\end{minipage}
	}
	\caption{The absolute value of the ratio $\mathcal{A}$ in the case of $Z^{(1)}_1\times Z^{(1)}_2 \to Z^{(2)}$. Each subfigure is labeled by the overtone numbers of the source modes in the format of $n_1 \times n_2$.
    In (a), (b), and (c), $\mathcal{A}$ diverge at the origin.
    In (a), there is also an isolated resonant point $(k_x, k_y)\approx(0.9864, 0.5388)$.
    The corresponding frequencies are $\omega^{(1)}_{n=0}=\omega^{(1)}_{m=0}=\frac{1}{2}\omega^{(1)}_{l=0}\approx 0.8862-0.1764i$.} 
	\label{SSS}
\end{figure}

\begin{figure}[hbpt]
	\centering
	\subfigure[]{
		\begin{minipage}[b]{.45\linewidth}
			\centering
			\includegraphics[scale=0.35]{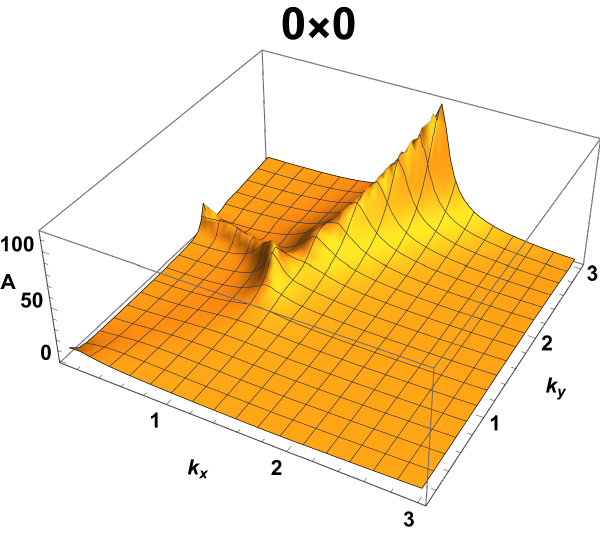}
		\end{minipage}
	}
	\subfigure[]{
		\begin{minipage}[b]{.45\linewidth}
			\centering
			\includegraphics[scale=0.35]{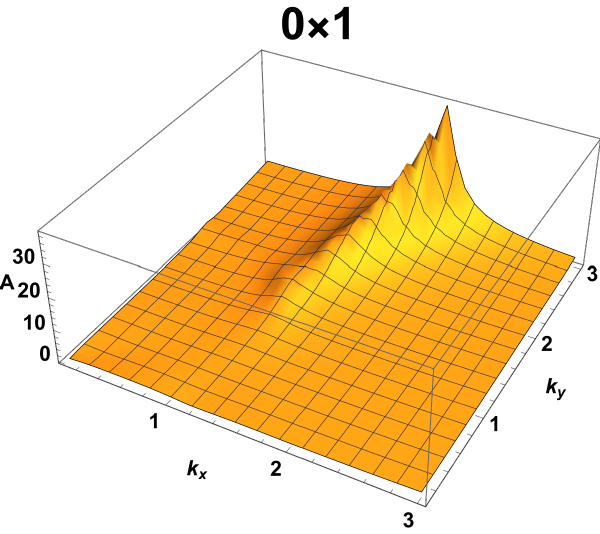}
		\end{minipage}
	}\\
	\subfigure[]{
		\begin{minipage}[b]{.45\linewidth}
			\centering
			\includegraphics[scale=0.35]{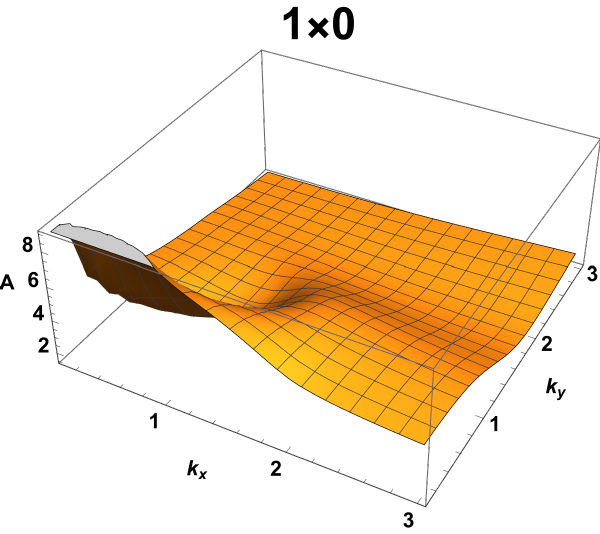}
		\end{minipage}
	}
	\subfigure[]{
		\begin{minipage}[b]{.45\linewidth}
			\centering
			\includegraphics[scale=0.35]{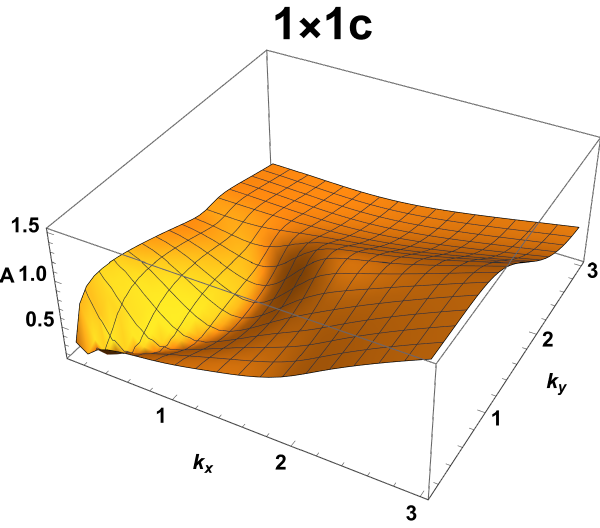}
		\end{minipage}
	}
	\caption{The absolute value of the ratio $\mathcal{A}$ in the case of $Y^{(1)}_1\times Z^{(1)}_2 \to Y^{(2)}$. Each subfigure is labeled by the overtone numbers of the source modes in the format of $n_1 \times n_2$.
    In (c), $\mathcal{A}$ diverge at the origin.} 
	\label{VSV}
\end{figure}

\begin{figure}[hbpt]
	\centering
	\subfigure[]{
		\begin{minipage}[b]{.45\linewidth}
			\centering
			\includegraphics[scale=0.35]{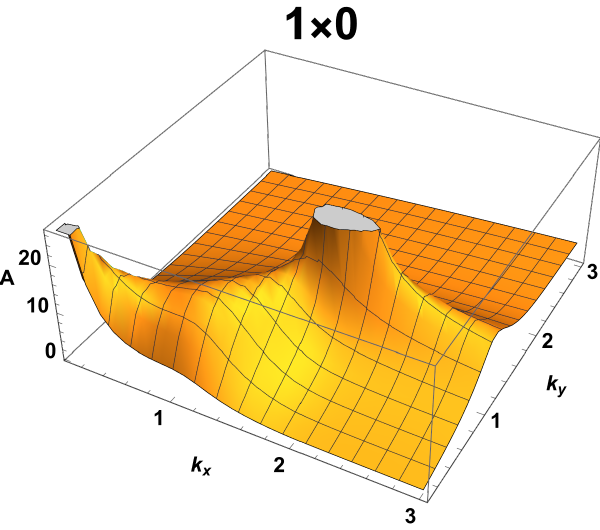}
		\end{minipage}
	}
	\subfigure[]{
		\begin{minipage}[b]{.45\linewidth}
			\centering
			\includegraphics[scale=0.35]{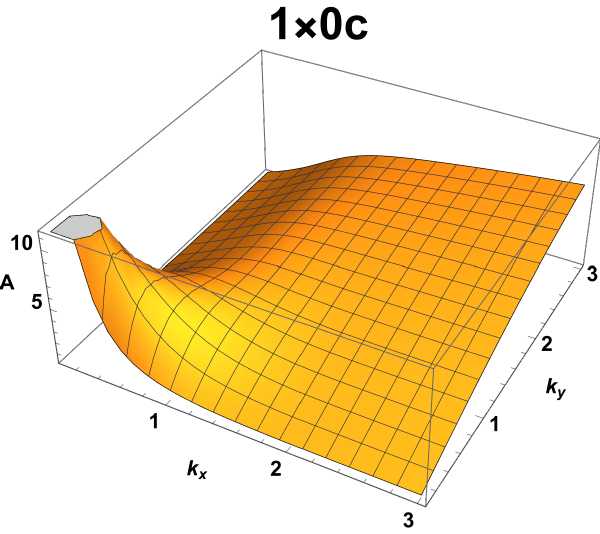}
		\end{minipage}
	}\\
	\subfigure[]{
		\begin{minipage}[b]{.45\linewidth}
			\centering
			\includegraphics[scale=0.35]{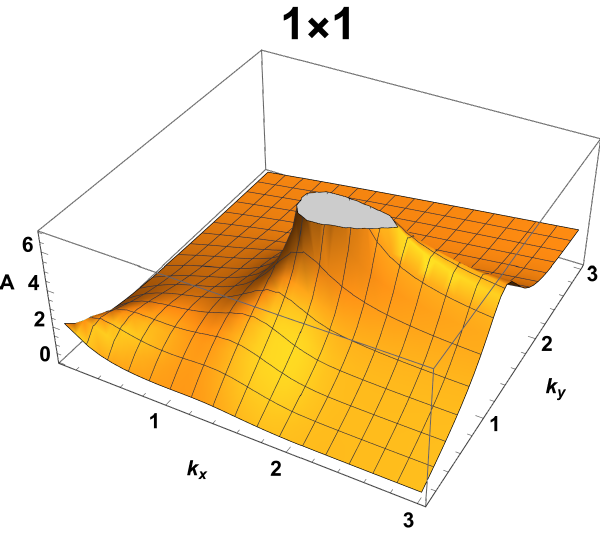}
		\end{minipage}
	}
	\subfigure[]{
		\begin{minipage}[b]{.45\linewidth}
			\centering
			\includegraphics[scale=0.35]{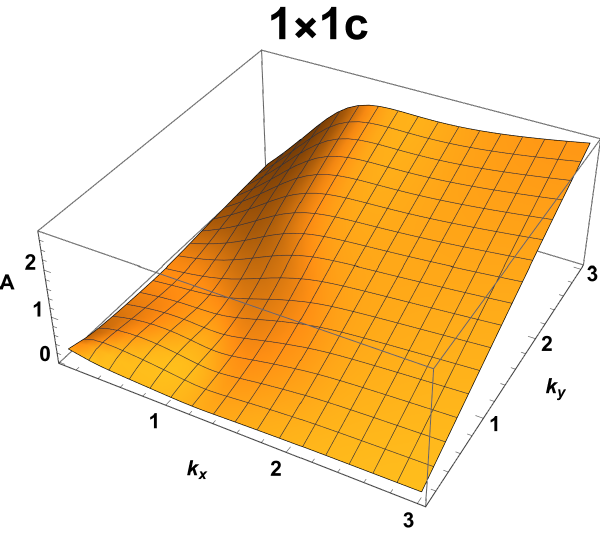}
		\end{minipage}
	}
	\caption{The absolute value of the ratio $\mathcal{A}$ in the case of $Y^{(1)}_1\times Z^{(1)}_2 \to Z^{(2)}$. Each subfigure is labeled by the overtone numbers of the source modes in the format of $n_1 \times n_2$.
    In (a) and (b), $\mathcal{A}$ diverge at the origin.
    In (a), the isolated resonant point locate at $(k_x, k_y)\approx (1.7793, 1.1923)$.
    The corresponding frequencies are $\omega^{(1)}_{n=1}\approx 2.1550-2.0400i$, $\omega^{(1)}_{m=0}\approx 1.9610-0.3727i$ and $\omega^{(1)}_{l=1}=\omega^{(1)}_{n=1}+\omega^{(1)}_{m=0}$.
    In (c), the isolated resonant point locate at $(k_x, k_y)\approx (1.6569, 1.2788)$.
    The corresponding frequencies are $\omega^{(1)}_{n=1}\approx 2.1095-2.0965i$, $\omega^{(1)}_{m=1}\approx 2.6812-2.5616i$ and $\omega^{(1)}_{l=2}=\omega^{(1)}_{n=1}+\omega^{(1)}_{m=1}$.} 
	\label{VSS}
\end{figure}

The amplitude ratio $\mathcal{A}$ for selected representative source modes of all the six processes shown in \eqref{process} is numerically evaluated and illustrated in Figures (\ref{VVV}), (\ref{VVS}), (\ref{SSV}), (\ref{SSS}), (\ref{VSV}), and (\ref{VSS}).
The labels on the subfigures correspond to the overtone numbers of two source modes.
Note that we set $k_{1x}=k_{2x}=k_x$ (i.e. $\vec{k}_1=(k_x,k_y)$ and $\vec{k}_2=(k_x,-k_y)$) in all figures so that the ratio depends only on $k_x$ and $k_y$ and can be conveniently plotted as a surface over the ($k_x$, $k_y$) plane. The location of the horizon $r_h$ is set to be $1$ for simplicity.

The figures reveal divergent amplitude ratios at special points in the $(k_x, k_y)$ plane, signaling the occurrence of resonance. 
Since the QNFs depend on the magnitude of wave vector, as two wave vectors of the source modes vary continuously, there exist special pairs of $(\vec{k}_1, \vec{k}_2)$ such that the combined frequency of two sources coincides with one of another linear-order quasi-normal mode, i.e.
\begin{equation}\label{res}	\omega^{(2)}=\omega^{(1)}_n(|\vec{k}_1|)+\omega^{(1)}_m(|\vec{k}_2|)=\omega^{(1)}_l(|\vec{k}_1+\vec{k}_2|),
\end{equation}
where $\omega^{(1)}_n$ and $\omega^{(1)}_m$ are the QNFs of the source modes, while $\omega^{(1)}_l$ is another QNF of vector or scalar modes (depending on the process under consideration). 
The indices $n$, $m$, and $l$ denote the overtone numbers of the modes.
N these resonant points, the ratio $\mathcal{A}$ is greatly amplified. 

We classify the resonant points into three forms, according to their location within the diagram and their topological characteristics, specifically whether they are isolated points or are connected to form continuous curves.
These features reflect different ways of how the resonant condition \eqref{res} is fulfilled and provide information on the structure of dispersion relations of vector and scalar modes. 
Specifically:

(1) The divergence at the original point of momentum space, as shown in Figure (\ref{VVV}a), (\ref{VVS}b), (\ref{SSV}b), (\ref{SSS}a), (\ref{SSS}b), (\ref{SSS}c), (\ref{VSS}b).
This form of divergence appears if one of the source modes is a hydrodynamic mode ($n=0$ or $m=0$). The resonance is located at $k_x=k_y=0$. 
This kind of divergences can be predicted according to conditions such as the isospectrality between vector and scalar modes at zero wave number.
For instance, the divergence observed at the origin in Figure (\ref{VVV}a) arises due to the vanishing frequency of hydrodynamic mode at zero wave vector, while the divergence at the origin in Figure (\ref{VVS}b) is the result of isospectrality between the vector and scalar sectors at zero wave vector.

(2) The divergence at isolated points except the origin, where the sum of the frequency of two source modes accidentally coincides with another QNF such that the resonance condition is fulfilled. This kind of pole is much more difficult to predict than the first form. 
The divergences of this form appear in Figure (\ref{VVS}c), (\ref{SSS}a), (\ref{VSS}a), (\ref{VSS}c).

(3) The divergent curve.
The character of the third form is that poles are connected to form continuous curves in momentum space. 
This form of divergences arises when two source modes are mirror modes, which have frequencies $\omega$ and $-\omega*$ respectively, and the sum of the frequencies coincides with the frequency of the vector hydrodynamic mode.
This form of divergences can be found in Figure (\ref{VVV}c), (\ref{SSV}a), (\ref{SSV}c).

\section{The holographic dual of the amplitude}\label{sec4}

Finally, we establish the holographic interpretation of the QQNM amplitudes by explicitly connecting them to fully retarded correlation functions in the boundary field theory. 
These correlation functions are defined as the expectation values of the R-product of operators \cite{Chou:1984es, Meltzer:2021bmb}:
\begin{small}
	\begin{equation}\label{Rproduct}
		\begin{aligned}
			G_R(x;x_1,...x_{n-1}):= (-i)^{n-1} \sum_{i} \theta\left(t-t_{i_{1}}\right) \theta\left(t_{i_{1}}-t_{i_{2}}\right) \ldots \theta\left(t_{i_{n-2}}-t_{i_{n-1}}\right)\\
			\left \langle {\left[\ldots\left[\left[O(x), O\left(x_{i_{1}}\right)\right], O\left(x_{i_{2}}\right)\right] \ldots, O\left(x_{i_{n-1}}\right)\right]}\right \rangle,
		\end{aligned}
	\end{equation} 
\end{small}
which reduces to the retarded two-point function for $n=2$.

The functions $A^{(1)}$ and $B^{(1)}$ in a linear-order ingoing solution \eqref{nearbdry} are holographically interpreted as the source and expectation value of a dual operator $O(x)$ of scaling dimension $\Delta$, respectively, where $\left \langle O(x)\right \rangle=(2\Delta-d)B^{(1)}(x)$.   
The two-point and three-point fully retarded correlation functions can be computed respectively as \cite{Pantelidou:2022ftm} 
\begin{equation}
	\begin{aligned}
		G_R(x-x')&=\left. \frac{\delta \left \langle O(x)\right \rangle}{\delta A^{(1)}(x')} \right|_{A^{(1)}=0},\\
		G_R(x;x',x'')&= \left. \frac{\delta^2 \left \langle O(x)\right \rangle}{\delta A^{(1)}(x') \delta A^{(1)}(x'')} \right|_{A^{(1)}=0}.
	\end{aligned}
\end{equation}
The three-point function vanishes if we only consider the first-order solutions, because $B^{(1)}$ is linear in $A^{(1)}$.
However, in the presence of a quadratic source such as \eqref{Source2nd} in the bulk, the expectation value is modified by a second-order solution $\psi^{(2)}=r^{-\Delta} B^{(2)}(x)$ which is quadratic in the source.
The two-point and three-point functions in momentum space are then found to be
	\begin{align}
		G_R(\omega,\vec{k})&=(2\Delta-d) \frac{B^{(1)}(\omega,\vec{k})}{A^{(1)}(\omega,\vec{k})},\label{Green2}\\
		G_R(\omega,\vec{k};\omega',\vec{k}')&=(2\Delta-d)\frac{B^{(2)}(\omega+\omega',\vec{k}+\vec{k}')}{A^{(1)}(\omega,\vec{k})A^{(1)}(\omega',\vec{k}')}.\label{Green3}
	\end{align}
We note that the QQNM amplitude ratio at the AdS boundary \eqref{QQNMA} is simply a ratio of $B^{(2)}$ and $B^{(1)}$.
Therefore, by combining \eqref{Green3} and \eqref{QQNMA}, the amplitude ratio is identified as the following ratio of correlation functions with both source frequencies set to be QNFs:
\begin{equation}\label{Amp}
	\mathcal{A}_{n\times m}=(2\Delta-d)\lim\limits_{
		\substack{\omega \to \omega_n \\ \omega' \to \omega_m} } \frac{G_R(\omega,\vec{k};\omega',\vec{k}')}{G_R(\omega,\vec{k})G_R(\omega',\vec{k}')}.
\end{equation}
where $\omega_n$ and $\omega_m$ denote specific QNFs.

The pole structure of the 3-point function can be analyzed using tree level Witten diagrams, as demonstrated in \cite{Pantelidou:2022ftm}. 
Depending on the number of legs corresponding to QNFs, the diagram exhibits single, double, or triple poles.
Divergences of the three-point function \eqref{Green3} occur at
\begin{equation}\label{WittenD}
   \omega=\omega_n(|\vec{k}|) \quad \text{or} \quad\omega'=\omega_n(|\vec{k}'|) \quad \text{or} \quad\omega+\omega'=\omega_n(|\vec{k}+\vec{k}'|), 
\end{equation} 
where $\omega_n(|\vec{k}|)$ denotes the dispersion relation of a QNM with overtone number $n$.
The first two conditions specified in equation \eqref{WittenD} describe cases that the external legs of the diagram are driven at QNFs. 
The final condition in \eqref{WittenD} represents a resonant excitation of a QNM, where the total incoming frequency matches a QNF, thereby resulting in an outgoing leg with QNF.
When one, two, or all three of these conditions are satisfied, corresponding to one, two, or three legs being associated with QNMs, the three-point function exhibits single, double, or triple poles, respectively.

In the calculation of QQNM amplitude ratios, we restrict our analysis to cases where two incoming legs correspond to QNM frequencies, as the case in equation \eqref{Amp} where the poles of incoming legs are canceled out with the poles of two-point functions.
However, if the third leg also corresponds to a QNF, both the three-point function and amplitude ratio become divergent. 
Therefore, the three-point function shares non-trivial poles with the amplitude ratio, located at momenta and frequencies satisfying the condition in \eqref{res}.
This connection is significant because the three-point function governs the system's nonlinear order response under external sources. Specifically, the nonlinear order response to a perturbation localized in time is determined by summing over the residues of $G_R(\omega,\vec{k};\omega',\vec{k}')$ at its poles, which is a direct extension of the case of linear response \cite{Amado:2007pv}.
This establishes a direct relationship between the amplitude ratio, the pole structure of correlation functions, and the dynamics of nonlinear order in a holographic system dual to the black brane.

\section{Discussion and Conclusion}\label{sec5}

In this work, we have investigated the quadratic QNMs in the AdS$_4$ black brane.
Such modes are driven by pairs of linear-order modes, with resonances emerging at specific wave vector values of the source modes. 
Such resonances are expected to be present in higher dimensional black branes because of their translational symmetry along the boundary directions. 
Inspired by the resonances discovered in the quadratic QNMs, we expect to explore the potential instabilities in the AdS black brane, just as the phenomenon observed in pure AdS spacetime \cite{Bizon:2011gg,Jalmuzna:2011qw}. Such exploration would provide new insights into the holographic understanding of nonlinear quantum phenomena such as quantum chaos and quantum turbulence.
Furthermore, moving beyond the ``frequency domain" approach and exploring the temporal evolution of the resonant modes by imposing initial conditions presents a compelling avenue for future investigation.

We have also found that the quadratic-order dynamics in the bulk governs the pole structure of fully retarded three-point correlation functions on the boundary.
This correspondence provides a concrete realization of the gauge/gravity duality, showing how the QQNM amplitudes in the bulk gravitational theory encode essential information about the dynamical response of the boundary quantum field theory.
It would also be interesting to explore nonlinear perturbations in important holographic setups for condensed matter systems, such as holographic superconductors \cite{Gubser:2008px,Hartnoll:2008vx} and holographic non-Fermi liquids\cite{Hartnoll:2009ns}.

\section*{Acknowledgement}

We thank Kai Li, Pan Li, Wei-Jia Li, Ya-Wen Sun, Yu Tian, and Hong-Bao Zhang for their valuable discussions.
This work is supported in part by the National Natural Science Foundation of China under Grant Nos. 12347183, 12035016, and 12275275.
It is also supported by Beijing Natural Science Foundation under Grant No. 1222031, and the Innovative Projects of Science and Technology No. E2545BU210 at IHEP.

\appendix

\section{The reconstruction of metric components}\label{A}

In this appendix, we reconstruct the metric components of vector type from the master variable $Y^{(1)}$.

Under an infinitesimal diffeomorphism $x^\mu \to x^\mu+\epsilon \xi^\mu$, where  
\begin{equation}
	\begin{aligned}
		\xi_\mu(t,x,y,r)=e^{-i \omega t+ i k_x x + i k_y y} \tilde{\xi}_{\mu}(\omega,k_x,k_y,r),
	\end{aligned}
\end{equation}
and
\begin{equation}
	\begin{aligned}
		\tilde{\xi}_{a}=&\zeta_{a}, \quad \quad \tilde{\xi}_{i}=&\zeta \hat{k}_{i}+\eta \hat{k}_{\perp i},
	\end{aligned}
\end{equation}
the first-order variables of the vector type transform as
\begin{equation}
	\begin{aligned}
		V^{(1)}_t \to &V^{(1)}_t-i \omega \frac{\eta}{r^2},\\
		V^{(1)}_r \to &V^{(1)}_r+ \left(\frac{\eta}{r^2}\right)',\\
		V^{(1)} \to &V^{(1)}+i k \frac{\eta}{r^2},
	\end{aligned}
\end{equation} 
where $k=\sqrt{k_x^2+k_y^2}$ and the prime denotes the derivative with respect to $r$.
It is straightforward to see that the following three combinations of variables 
\begin{equation}\label{gaugeInv}
	\begin{aligned}
		E_1&=k V^{(1)}_t+\omega V^{(1)},\\
		E_2&=(V^{(1)}_t)' +i\omega V^{(1)}_r,\\
		E_3&=(V^{(1)})'-i k V^{(1)}_r,
	\end{aligned}
\end{equation}
are invariant under the diffeomorphism. 
The Einstein equations of the vector sector consist of three equations which can be represented fully by three gauge invariants as
\begin{equation}\label{EinEq}
	\begin{aligned}
		-\frac{1}{2} f(r) (r^4 E_2)'+\frac{k}{2} E_1=0,\\
		\frac{i\omega}{2 f(r)}E_2+\frac{ik}{2} E_3=0,\\
		-\frac{1}{2} (r^4 f(r) E_3)'-\frac{\omega}{2 f(r)}E_1=0.
	\end{aligned}
\end{equation}
These gauge invariant variables are not independent and obey the constraint
\begin{equation}\label{constr}
	E_1'-k E_2-\omega E_3 =0.
\end{equation}
Following \cite{Kovtun:2005ev}, we choose $Y^{(1)}:=\frac{E_1}{2\pi T}$ to be our master variable of the vector sector.
$Y^{(1)}$ satisfies the second-order ODE \eqref{Master1stY} which can be easily derived from \eqref{EinEq} and \eqref{constr}.

In order to reconstruct the metric components from the master variable, we choose the Regge-Wheeler gauge, which imposes $V^{(1)}=0$ for vector type variables.
After fixing the gauge, we have $V_t=\frac{E_1}{k}$ and $V_r=\frac{i E_3}{k}$.
Then, by combining the second equation in \eqref{EinEq} and \eqref{constr}, we obtain $E_3=\frac{\omega}{\omega^2-k^2 f(r)}E_1'$.
Therefore we have
\begin{equation}
	\begin{aligned}
		V^{(1)}_{t}=&\frac{E_1}{k}, \quad \quad V^{(1)}_{r}=&\frac{i \omega}{k(\omega^2-k^2 f(r))}E_1'.
	\end{aligned}
\end{equation}
After a normalization with $2\pi T$, we obtain \eqref{reconst} and the reconstruction for vector type variables is completed.

It is worthwhile to note that the same gauge fixing can be applied to both modes in the source simultaneously.
To impose $V^{(1)}=0$, we need to set $\eta=\frac{i r^2 V^{(1)}}{k}$.
Writing the dependence of $\omega$ and $\vec{k}$ in the variables explicitly, we have
\begin{equation}
	\eta(\omega, \vec{k}, r)=\frac{i r^2 }{k} V^{(1)}(\omega, \vec{k}, r).
\end{equation}
Since $\eta(\omega_1, \vec{k_1}, r)$ and $\eta(\omega_2, \vec{k_2}, r)$ are generally independent, they can be used to eliminate the component $V^{(1)}$ in each of two source modes (i.e., $V^{(1)}(\omega_1, \vec{k_1}, r)$ and $V^{(1)}(\omega_2, \vec{k_2}, r)$), respectively.

\section{The coefficients $F_j$}\label{B}
\begin{itemize}
	\item For the process $Y^{(1)}_1\times Y^{(1)}_2 \to Y^{(2)}$
	\begin{small}
		\begin{equation}
			\begin{aligned}
				F^{YYY}_1=&(2 k_y ((-1 + 
				u^3) (-((k_{2x}^2 + k_y^2) (k_{1x}^3 + k_{1x}^2 k_{2x} + k_{2x}^3 + 
				k_{1x} (2 k_{2x}^2 + k_y^2)) \omega_1) \\
				&+ (k_{1x}^2 + 
				k_y^2) (k_{1x}^3 + 2 k_{1x}^2 k_{2x} + k_{1x} k_{2x}^2 + k_{2x}^3 + k_{2x} k_y^2) \omega_2)+ \omega_1 \omega_2 (k_{1x}^3 \omega_1 \\
				&+ k_{1x}^2 k_{2x} (\omega_1 - 2 \omega_2) + 
				k_{1x} k_{2x}^2 (2 \omega_1 - \omega_2) - k_{2x}^3 \omega_2 + 
				k_{1x} k_y^2 (\omega_1 + \omega_2) \\
				&-k_{2x} k_y^2 (\omega_1 + \omega_2))))/((k_{1x}^2 + k_y^2) (k_{2x}^2 + 
				k_y^2) r_h^2 ((k_{1x}^2 + k_y^2) (-1 + u^3) \\
				&+ \omega_1^2) ((k_{2x}^2 + 
				k_y^2) (-1 + u^3) + \omega_2^2))
			\end{aligned}        
		\end{equation}

		\begin{equation}
			\begin{aligned}
				F^{YYY}_2=&(3 (k_{1x} + 
				k_{2x}) k_y u^2 ((2 k_{1x}^4 + 4 k_{1x}^3 k_{2x} + 4 k_{1x} k_{2x}^3 + k_{2x}^4 + k_y^4 \\
				&+ 
				2 k_{1x}^2 (3 k_{2x}^2 + k_y^2)) \omega_1 + (k_{1x}^2 + k_y^2)^2 \omega_2))/((k_{1x}^2 + k_y^2) (k_{2x}^2 + 
				k_y^2) r_h^2 ((k_{1x}^2 \\
				&+ k_y^2) (-1 + u^3) + \omega_1^2) ((k_{1x} + 
				k_{2x})^2 (-1 + u^3) + (\omega_1 + \omega_2)^2))
			\end{aligned}
		\end{equation}
		
		\begin{equation}
			\begin{aligned}
				F^{YYY}_3=&(3 (k_{1x} + k_{2x}) k_y u^2 (-(k_{2x}^2 + k_y^2)^2 \omega_1 - 
				k_{1x}^4 \omega_2 - (2 k_{2x} (k_{1x} + k_{2x}) (2 k_{1x}^2 \\
				&+ k_{1x} k_{2x} + k_{2x}^2) + 
				2 k_{2x}^2 k_y^2 + k_y^4) \omega_2))/((k_{1x}^2 + k_y^2) (k_{2x}^2 + 
				k_y^2) r_h^2 ((k_{2x}^2 \\
				&+ k_y^2) (-1 + u^3) + \omega_2^2) ((k_{1x} + 
				k_{2x})^2 (-1 + u^3) + (\omega_1 + \omega_2)^2))
			\end{aligned}
		\end{equation}
		
		\begin{equation}
			\begin{aligned}
				F^{YYY}_4=&(2 k_y (k_{1x}^3 \omega_1 + k_{1x}^2 k_{2x} (\omega_1 - 2 \omega_2) + 
				k_{1x} k_{2x}^2 (2 \omega_1 - \omega_2) - k_{2x}^3 \omega_2 \\
				&+ 
				k_{1x} k_y^2 (\omega_1 + \omega_2) - 
				k_{2x} k_y^2 (\omega_1 + \omega_2)))/((k_{1x}^2 + k_y^2) (k_{2x}^2 + 
				k_y^2) r_h^4 (-1 + u^3)^2)
			\end{aligned}
		\end{equation}
	\end{small}
	
	\item For the process $Y^{(1)}_1\times Y^{(1)}_2 \to Z^{(2)}$
	\begin{small}
		\begin{equation}
			\begin{aligned}
				F^{YYZ}_1=&-(((k_{1x} + k_{2x})^2 (3 k_{1x} k_{2x} - 2 k_y^2) (k_{1x}^2 + k_y^2) (k_{2x}^2 + 
				k_y^2) (-1 + u^3)^2 \\
				&- (1 - 
				u^3) (k_{1x}^5 k_{2x} (3 k_{2x}^2 + 3 k_y^2 + 4 \omega_2^2) + k_{1x}^3 k_{2x} (3 k_{2x}^4 + 3 k_y^4 - 4 k_y^2 \omega_1 \omega_2 \\
				&+ 
				6 k_{2x}^2 (k_y^2 - \omega_1 \omega_2)) + 
				k_{1x} (k_{2x}^5 (3 k_y^2 + 4 \omega_1^2) + 
				k_{2x}^3 (3 k_y^4 - 4 k_y^2 \omega_1 \omega_2) \\
				&- 
				2 k_{2x} k_y^4 (2 \omega_1^2 - 3 \omega_1 \omega_2 + 
				2 \omega_2^2)) + 
				k_{1x}^2 (k_{2x}^4 (6 k_y^2 + 6 \omega_1^2 - \omega_1 \omega_2) \\
				&+ 
				k_y^4 (-2 \omega_1^2 + \omega_1 \omega_2 - 
				4 \omega_2^2) + 
				2 k_{2x}^2 k_y^2 (3 k_y^2 + 2 \omega_1^2 - 
				3 \omega_1 \omega_2 + 2 \omega_2^2)) \\
				&- 
				k_y^2 (k_{2x}^4 \omega_1 (2 \omega_1 - \omega_2) + 
				2 k_y^4 (\omega_1 + \omega_2)^2 + 
				k_{2x}^2 k_y^2 (4 \omega_1^2 - \omega_1 \omega_2 + 
				2 \omega_2^2)) \\
				&+ 
				k_{1x}^4 (6 k_{2x}^4 + k_y^2 (\omega_1 - 2 \omega_2) \omega_2 + 
				k_{2x}^2 (6 k_y^2 - \omega_1 \omega_2 + 
				6 \omega_2^2))) \\
				&+ \omega_1 \omega_2 (3 k_{1x}^4 (k_{2x}^2 -
				k_y^2) + 
				2 k_{1x}^3 k_{2x} (3 k_{2x}^2 - 6 k_y^2 + 2 \omega_1 \omega_2) + 
				2 k_{1x} k_{2x} (-3 k_y^4 \\
				&+ 
				k_{2x}^2 (-6 k_y^2 + 2 \omega_1 \omega_2) + 
				4 k_y^2 (\omega_1^2 + \omega_1 \omega_2 + \omega_2^2)) +
				k_{1x}^2 (3 k_{2x}^4 - 3 k_y^4 \\
				&+ 
				4 k_y^2 (\omega_1^2 + \omega_1 \omega_2 + \omega_2^2) - 
				2 k_{2x}^2 (9 k_y^2 + 2 (\omega_1^2 + \omega_2^2))) + 
				k_y^2 (-3 k_{2x}^4 \\
				&+ 4 k_y^2 (\omega_1 + \omega_2)^2 + 
				k_{2x}^2 (-3 k_y^2 + 
				4 (\omega_1^2 + \omega_1 \omega_2 +
				\omega_2^2)))))/((k_{1x}^2 + k_y^2) \\
				&(k_{2x}^2 + 
				k_y^2) r_h^2 ((k_{1x}^2 + k_y^2) (-1 + u^3) + \omega_1^2) ((k_{2x}^2 + 
				k_y^2) (-1 + u^3) + \omega_2^2)))
			\end{aligned}        
		\end{equation}

		\begin{equation}
			\begin{aligned}
				F^{YYZ}_2=&(6 (k_{1x} + k_{2x}) u^2 (k_{1x}^6 k_{2x} (2 + u^3) + 
				k_{1x}^5 (-k_y^2 (-1 + u^3) + 3 k_{2x}^2 (1 + u^3)) \\
				&+ 
				k_{1x}^3 (k_{2x}^4 (-1 + u^3) - k_y^4 (-1 + u^3) + 
				k_{2x}^2 (12 k_y^2 + \omega_1 (5 \omega_1 - \omega_2)) \\
				&+ 
				k_y^2 \omega_1 (-\omega_1 + \omega_2)) + 
				k_{1x} (k_{2x}^4 (k_y^2 (-1 + u^3) + 2 \omega_1^2) - 
				k_{2x}^2 k_y^2 (3 k_y^2 (-3 + u^3) \\
				&+ \omega_1 (\omega_1 - 
				5 \omega_2)) - 
				k_y^4 \omega_1 (\omega_1 + \omega_2)) + 
				k_{1x}^4 k_{2x} (3 k_{2x}^2 u^3 - 2 k_y^2 (-4 + u^3) \\
				&- 
				2 (\omega_1^2 + 4 \omega_1 \omega_2 + 2 \omega_2^2)) - 
				k_{2x} k_y^2 (k_{2x}^2 (k_y^2 (-4 + 
				u^3) + \omega_1 (\omega_1 - \omega_2)) \\
				&+ 
				k_y^2 (5 \omega_1^2 + 9 \omega_1 \omega_2 + 
				4 \omega_2^2)) + 
				k_{1x}^2 (k_{2x}^3 (2 k_y^2 (2 + 
				u^3) + \omega_1 (5 \omega_1 - \omega_2))\\
				& - 
				k_{2x} k_y^2 (3 k_y^2 (-2 + u^3) + 9 \omega_1^2 + 
				11 \omega_1 \omega_2 + 8 \omega_2^2))))/\left((k_{1x}^2 + 
				k_y^2) (k_{2x}^2 + k_y^2) \right.\\
				& \left.r_h^2 (k_{1x}^2 (-1 + u^3)+ 
				k_y^2 (-1 + u^3) + \omega_1^2) (k_{1x}^2 (-4 + u^3) + 
				2 k_{1x} k_{2x} (-4 + u^3) \right.\\
				& \left.+ k_{2x}^2 (-4 + u^3) + 
				4 (\omega_1 + \omega_2)^2)\right)
			\end{aligned}
		\end{equation}
		
		\begin{equation}
			\begin{aligned}
				F^{YYZ}_3=&(6 (k_{1x} + 
				k_{2x}) u^2 (k_{1x}^4 k_{2x} (k_{2x}^2 (-1 + u^3) + k_y^2 (-1 + u^3) + 
				2 \omega_2^2) + 
				k_{1x} (k_{2x}^6 (2 + u^3) \\
				&- 
				2 k_{2x}^4 (k_y^2 (-4 + u^3) + 2 \omega_1^2 + 
				4 \omega_1 \omega_2 + \omega_2^2) - 
				k_y^4 (4 \omega_1^2 + 9 \omega_1 \omega_2 + 5 \omega_2^2) \\
				&-
				k_{2x}^2 k_y^2 (3 k_y^2 (-2 + u^3) + 8 \omega_1^2 + 
				11 \omega_1 \omega_2 + 9 \omega_2^2))- 
				k_{2x} k_y^2 (k_{2x}^4 (-1 + u^3)\\
				&+ 
				k_y^2 \omega_2 (\omega_1 + \omega_2) + 
				k_{2x}^2 (k_y^2 (-1 + 
				u^3) + \omega_2 (-\omega_1 + \omega_2))) + 
				k_{1x}^2 (3 k_{2x}^5 (1 + u^3) \\
				&- 
				k_{2x} k_y^2 (3 k_y^2 (-3 + 
				u^3) + \omega_2 (-5 \omega_1 + \omega_2)) + 
				k_{2x}^3 (12 k_y^2 + \omega_2 (-\omega_1 + 5 \omega_2))) \\
				&+ 
				k_{1x}^3 (3 k_{2x}^4 u^3 - 
				k_y^2 (k_y^2 (-4 + u^3) + \omega_2 (-\omega_1 + \omega_2))+ 
				k_{2x}^2 (2 k_y^2 (2 + u^3) \\
				&+ \omega_2 (-\omega_1 + 
				5 \omega_2)))))/((k_{1x}^2 + k_y^2) (k_{2x}^2 + 
				k_y^2) r_h^2 (k_{2x}^2 (-1 + u^3) + 
				k_y^2 (-1 + u^3) \\
				&+ \omega_2^2) (k_{1x}^2 (-4 + u^3)+ 
				2 k_{1x} k_{2x} (-4 + u^3) + k_{2x}^2 (-4 + u^3) + 
				4 (\omega_1 + \omega_2)^2))
			\end{aligned}
		\end{equation}
		
		\begin{equation}
			\begin{aligned}
				F^{YYZ}_4=&-((k_{1x}^6 (-4 + u^3) (3 k_{2x}^2 u^3 + k_y^2 (-4 + u^3)) + 
				2 k_{1x}^5 k_{2x} (-9 r_h^2 u^4 + 9 r_h^2 u^7 \\
				&+ 6 k_{2x}^2 u^3 (-4 + u^3) + 
				k_y^2 (40 - 14 u^3 + u^6) - 8 \omega_1 \omega_2 + 
				2 u^3 \omega_1 \omega_2) \\
				&+ 
				k_y^2 (k_{2x}^6 (-4 + u^3)^2 + 16 k_y^2 (\omega_1 + \omega_2)^4 + 
				8 k_{2x}^2 (\omega_1 + \omega_2)^2 (k_y^2 (-4 + u^3) \\
				&+ 
				2 (\omega_1^2 + \omega_1 \omega_2 + \omega_2^2)) + 
				k_{2x}^4 (k_y^2 (-4 + u^3)^2 - 18 r_h^2 u^4 (-1 + u^3) \\
				&+ 
				4 (-4 + u^3) (2 \omega_1^2 + 3 \omega_1 \omega_2 + 
				2 \omega_2^2))) + 
				4 k_{1x}^3 k_{2x} (3 k_{2x}^4 u^3 (-4 + u^3)\\
				&+ k_y^4 (-4 + u^3)^2 + 
				4 \omega_1 \omega_2 (\omega_1 + \omega_2)^2 + 
				k_{2x}^2 (27 r_h^2 u^4 (-1 + u^3) - k_y^2 (-56 \\
				&+ 10 u^3 + u^6) + 
				8 \omega_1^2 + 4 u^3 \omega_1^2 - 
				8 \omega_1 \omega_2+ 14 u^3 \omega_1 \omega_2 + 
				8 \omega_2^2 + 4 u^3 \omega_2^2) \\
				&- 
				2 k_y^2 (9 r_h^2 u^4 (-1 + u^3) - 
				2 ((-7 + u^3) \omega_1^2 + (-10 + 
				u^3) \omega_1 \omega_2 + (-7 + 
				u^3) \omega_2^2))) \\
				&+ k_{1x}^4 (18 k_{2x}^4 u^3 (-4 + u^3) + 
				k_y^2 (k_y^2 (-4 + u^3)^2 - 18 r_h^2 u^4 (-1 + u^3) \\
				&+ 
				4 (-4 + u^3) (2 \omega_1^2 + 3 \omega_1 \omega_2 + 
				2 \omega_2^2)) + k_{2x}^2 (-k_y^2 (-176 + 40 u^3 + u^6) \\
				&+ 
				8 (9 r_h^2 u^4 (-1 + u^3) + (2 + u^3) \omega_1^2 + 
				4 (-1 + u^3) \omega_1 \omega_2 + (2 + 
				u^3) \omega_2^2))) \\
				&+ k_{1x}^2 (3 k_{2x}^6 u^3 (-4 + u^3) + 
				8 k_y^2 (\omega_1 + \omega_2)^2 (k_y^2 (-4 + u^3) + 
				2 (\omega_1^2 + \omega_1 \omega_2 + \omega_2^2)) \\
				&+ 
				2 k_{2x}^2 (3 k_y^4 (-4 + u^3)^2 - 
				8 (\omega_1 + \omega_2)^2 (\omega_1^2 + \omega_2^2) + 
				k_y^2 (-54 r_h^2 u^4 (-1 + u^3) \\
				&+ 
				8 (-10 + u^3) \omega_1^2 + 
				4 (-28 + u^3) \omega_1 \omega_2 + 
				8 (-10 + u^3) \omega_2^2)) + 
				k_{2x}^4 (-k_y^2 (-176 \\
				&+ 40 u^3 + u^6) + 
				8 (9 r_h^2 u^4 (-1 + u^3) + (2 + u^3) \omega_1^2 + 
				4 (-1 + u^3) \omega_1 \omega_2 \\
				&+ (2 + 
				u^3) \omega_2^2))) + 
				2 k_{1x} (k_{2x}^5 (9 r_h^2 u^4 (-1 + u^3) + k_y^2 (40 - 14 u^3 + u^6) \\
				&+ 
				2 (-4 + u^3) \omega_1 \omega_2) + 
				8 k_{2x} k_y^2 (\omega_1 + \omega_2)^2 (k_y^2 (-4 + u^3) + 
				2 (\omega_1^2 + \omega_1 \omega_2 + \omega_2^2)) \\
				&+ 
				2 k_{2x}^3 (k_y^4 (-4 + u^3)^2 + 
				4 \omega_1 \omega_2 (\omega_1 + \omega_2)^2 - 
				2 k_y^2 (9 r_h^2 u^4 (-1 + u^3)- 
				2 ((-7 \\
				&+ u^3) \omega_1^2 + (-10 + 
				u^3) \omega_1 \omega_2 + (-7 + 
				u^3) \omega_2^2)))))/((k_{1x}^2 + k_y^2) (k_{2x}^2 + 
				k_y^2) \\
				&r_h^4 (-1 + u^3)^2 (k_{1x}^2 (-4 + u^3) + 
				2 k_{1x} k_{2x} (-4 + u^3) + k_{2x}^2 (-4 + u^3) \\
				&+ 
				4 (\omega_1 + \omega_2)^2)))
			\end{aligned}
		\end{equation}
	\end{small}		
\end{itemize}

\section{Pseudo-spectral method}\label{C}
In this appendix, we concisely introduce the pseudo-spectral method \cite{Grandclement:2007sb,Monteiro:2009ke,Jansen:2017oag} for solving quasi-normal modes and quadratic modes. We illustrate this with the vector mode $Y$ as an example. The equation governing the first-order vector mode $Y^{(1)}$ is given by \eqref{Master1stY}. We first introduce a coordinate transformation $u=\frac{r_h}{r}$, which maps the AdS boundary to $u=0$ and the horizon to $u=1$. Without loss of generality we set $r_h$ to unity here; thus, $\mathfrak{w} = \frac{2\omega}{3}$ and $\mathfrak{q} = \frac{2 k}{3}$. Consequently, the equation transforms into
\begin{footnotesize}
	\begin{align}
		\quad(Y^{(1)})''(u)-\frac{\left(2 k^2 \left(u^3-1\right)^2-\left(u^3+2\right) \omega ^2\right) }{u \left(u^3-1\right)\left(k^2 \left(u^3-1\right)+\omega ^2\right)}(Y^{(1)})'(u)+\frac{ \left(k^2 \left(u^3-1\right)+\omega^2\right)}{\left(u^3-1\right)^2}Y^{(1)}(u)=0.
	\end{align}
\end{footnotesize}

Now we analyze the behavior near the horizon and near the boundary to match the boundary conditions. Starting with the AdS boundary, by plugging in an ansatz $Y^{(1)}(u)=u^p$, we find that there are two solutions, a non-normalizable mode $Y^{(1)}(u) = \text{const} + \mathcal{O}(u)$ which we must discard, and a normalizable mode $Y^{(1)}(u) = u^3 + \mathcal{O}(u^4)$. If we define $Y^{(1)}(u) = u^3 \widetilde{Y}^{(1)}(u)$, then the normalizable mode is smooth near the boundary, while the non-normalizable mode diverges.

Again, by plugging in an ansatz $Y^{(1)}(u)=(1-u)^p$, we find that two independent solutions are $Y^{(1)}(u)=(1-u)^{\frac{i \omega}{3}}\left(1+\mathcal{O}(1-u)\right)$ and $Y^{(1)}(u)=(1-u)^{-\frac{i \omega}{3}}\left(1+\mathcal{O}(1-u)\right)$. The former solution corresponds to an in-going wave, while the latter solution corresponds to an out-going wave. If we define $Y^{(1)}(u) = (1-u)^{-\frac{i \omega}{3}}  \overline{Y}^{(1)}(u)$, the in-going mode is smooth near the horizon, while the out-going mode, which oscillates more and more rapidly as it approaches the horizon, should be discarded.

Thus, by applying transformation $Y^{(1)}(u)= u^3 (1-u)^{-\frac{i \omega}{3}} y^{(1)}(u)$, the equation becomes
\begin{align}
	a(\omega,k) y^{(1)''}(u) + b(\omega,k) y^{(1)'}(u) + c(\omega,k) y^{(1)}(u)=0,
\end{align}
where 
\begin{small}
	\begin{align}
		a(\omega,k)&=9 (u-1) u \left(u^2+u+1\right)^2 \left(k^2 \left(u^3-1\right)+\omega ^2\right),\notag\\
		b(\omega,k)&=-3 i \left(u^2+u+1\right)2 k^2 (u-1) \left(u^2+u+1\right)^2 (u (\omega +6 i)-6 i)\nonumber\notag\\
		&\quad-3 i \left(u^2+u+1\right)\omega ^2 \left(u^3 (2 \omega +21 i)+2 u^2 \omega +2 u \omega -12 i\right),\notag\\
		c(\omega,k)&=9 k^4 (u-1) u \left(u^2+u+1\right)^2-k^2 \left(u^2+u+1\right)\omega  \left(u^5 (\omega +9 i) +2 u^4 (\omega +3 i)\right.\notag\\
		&\left. \quad+3 u^3(\omega +i)+2 u^2 (\omega -9 i)-u (17 \omega +15 i)-12i\right)-\omega ^2 \left(u^4 (\omega +9 i)^2\right.\notag\\
		&\quad\left.+3 u^3 \left(\omega^2+11 i \omega -27\right)+u^2 \left(6 \omega ^2+45 i \omega-81\right)+u \omega  (8 \omega +27 i)+12 i \omega \right),\notag
	\end{align}		
\end{small}
and the Dirichlet boundary condition requires that $y^{(1)}(u)$ is smooth at horizon and AdS boundary.

Next, we discretize the differential equations. Specifically, we substitute the continuous variables with a discrete collection of collocation points known as the grid points, and we expand the functions using specific basis functions referred to as cardinal functions. Typically, Chebyshev grids and Lagrange cardinal functions are employed:
\begin{align}
	x_{i}    =\cos \left(\frac{i}{N} \pi\right),\quad                                     
	C_{j}(x) =\prod_{i=0, i \neq j}^{N} \frac{x-x_{i}}{x_{j}-x_{i}}, i=0, \ldots, N .
\end{align}
Subsequently, the ordinary differential equation is transformed into a matrix equation
\begin{align}
	\left(M_0 + M_1 \omega + M_2 \omega^2 + \ldots + M_p \omega^p \right) y^{(1)} = 0,
\end{align}
where $y^{(1)}$ denotes the coefficients vector of $y^{(1)}(u)$, and $M_i (i=0,1,\ldots,p)$, which are purely numerical matrices, denote the linear combination of the derivative matrices. By defining
\begin{equation}
	M_{0}  =\left(\begin{array}{ccccc}
		M_0    & M_1    & M_2    & \cdots & M_{p-1} \\
		0      & 1      & 0      & \cdots & 0       \\
		0      & 0      & 1      & \cdots & 0       \\
		\vdots & \vdots & \vdots & \ddots & \vdots  \\
		0      & 0      & 0      & 0      & 1
	\end{array}\right),\quad 
	M_{1}  =\left(\begin{array}{cccccc}
		0      & 0      & 0      & \cdots & 0      & M_{p}  \\
		-1     & 0      & 0      & \cdots & 0      & 0      \\
		0      & -1     & 0      & \cdots & 0      & 0      \\
		\vdots & \vdots & \vdots & \ddots & \vdots & \vdots \\
		0      & 0      & 0      & \cdots & -1     & 0
	\end{array}\right),  
\end{equation}
we can rewrite this matrix equation as a generalized eigenvalue equation $(\tilde{M}_0 + \omega \tilde{M}_1)\tilde{y} = 0$, where $\tilde{y} = (y^{(1)}, \omega y^{(1)}, \omega^2 y^{(1)},\ldots, \omega^{p-1} y^{(1)})$. The linear quasi-normal frequency $\omega$ can be obtained by solving this generalized eigenvalue equation.

Quadratic modes satisfy the same equation as linear modes, except with an additional source term:
\begin{align}
	a(\omega^{(2)},k^{(2)}) y^{(2)''}(u) + b(\omega^{(2)},k^{(2)}) y^{(2)'}(u)
	+ c(\omega^{(2)},k^{(2)}) y^{(2)}(u)\nonumber\\
	=S\left[y^{(1)}_1\left(\omega^{(1)}_1,\vec{k}^{(1)}_1\right),y^{(1)}_2\left(\omega^{(1)}_2,\vec{k}^{(1)}_2\right)\right],
\end{align}
where $\omega^{(2)} = \omega^{(1)}_1 + \omega^{(1)}_2$ and $k^{(2)}=\left|{\vec{k}^{(1)}_1+\vec{k}^{(1)}_2}\right|$ are frequency and magnitude of quadratic mode, respectively. Therefore, after discretization, $y^{(2)}$ satisfies a linear equation
\begin{align}
	\left(M_0 + M_1 \omega^{(2)} + M_2 (\omega^{(2)})^2 + \ldots + M_p (\omega^{(2)})^p \right) y^{(2)} = s,
\end{align}
where $s$ denotes the coefficients vector of the source. The quadratic mode can be obtained by solving this linear equation.

\section{The tensor modes in AdS$_5$}\label{D}
In this appendix, we discuss the QQNM amplitudes in the case of the black brane in AdS$_5$. 
We focus on the tensor modes that are absent in AdS$_4$ and only consider the processes in which two tensor modes serve as the source.
The QQNM amplitudes become more difficult to converge in the case of AdS$_5$ in our numerical calculation of the H-H method, as it usually takes $N>45$. 
But if our goal is to find the pole, it is enough to check at which points in the momentum space the resonant condition \eqref{res} is satisfied.

In AdS$_5$, the metric fluctuation can be decomposed as 
\begin{equation}
    \begin{aligned}
		\tilde{h}^{(n)}_{ab}=&H^{(n)}_{ab},\\
		\tilde{h}^{(n)}_{ai}=&  H^{(n)}_{a}\hat{k}_{i}+  V^{(n)}_{a}\hat{k}^{(1)}_{\perp i}+\bar{V}^{(n)}_{a}\hat{k}^{(2)}_{\perp i},\\
		\tilde{h}^{(n)}_{ij}=& H^{(n)}_{L} \hat{k}_{i}\hat{k}_{j}+  H^{(n)}_{T}(\hat{k}^{(1)}_{\perp i}\hat{k}^{(1)}_{\perp j}+\hat{k}^{(2)}_{\perp i}\hat{k}^{(2)}_{\perp j})\\&+  V^{(n)}(\hat{k}_{i}\hat{k}^{(1)}_{\perp j}+\hat{k}_{j}\hat{k}^{(1)}_{\perp i})+ \bar{V}^{(n)}(\hat{k}_{i}\hat{k}^{(2)}_{\perp j}+\hat{k}_{j}\hat{k}^{(2)}_{\perp i})
        \\&+T^{(n)}(\hat{k}^{(1)}_{\perp i}\hat{k}^{(1)}_{\perp j}-\hat{k}^{(2)}_{\perp i}\hat{k}^{(2)}_{\perp j})+\bar{T}^{(n)}(\hat{k}^{(1)}_{\perp i}\hat{k}^{(2)}_{\perp j}+\hat{k}^{(2)}_{\perp i}\hat{k}^{(1)}_{\perp j}),		
	\end{aligned}
\end{equation}
where $\hat{k}$, $\hat{k}^{(1)}_{\perp}$, and $\hat{k}^{(2)}_{\perp}$ are orthonormal bases in three dimensional momentum space. 
They satisfy $\delta_{ij}=\hat{k}_i \hat{k}_j+\hat{k}^{(1)}_{\perp i}\hat{k}^{(1)}_{\perp j}+\hat{k}^{(2)}_{\perp i}\hat{k}^{(2)}_{\perp j}$.
There are two tensor modes, two vector modes, and one scalar mode. 
\begin{equation}
	\begin{aligned}
            &\text{ Tensor 1:}  &T^{(n)}; \quad &\\
            &\text{ Tensor 2:}  &\bar{T}^{(n)}; \quad &\\
		&\text{ Vector 1:}  &V^{(n)}_{a}, \quad & V^{(n)};\\
            &\text{ Vector 2:}  &\bar{V}^{(n)}_{a}, \quad & \bar{V}^{(n)};\\
		&\text{ Scalar:}  &H^{(n)}_{ab}, \quad & H^{(n)}_{a},  & H^{(n)}_{L}, \quad & H^{(n)}_{T}.
	\end{aligned}
\end{equation}
The QNM spectra for two tensors are identical, so is true for the spectra of two vectors.
There are more possible combinations for the coupling in source.
For instance, the process $\text{Tensor} \times \text{Tensor} \rightarrow \text{Vector}$ now consists of twelve sub-processes such as
$\text{Tensor 1} \times \text{Tensor 2} \rightarrow \text{Vector 2}$.
The QQNM amplitudes of these sub-processes may differ with one another. 
However, the resonant poles should be the same for all such sub-processes, since the spectrum is the same in each sector.
In order to find the divergences, it is enough to search for resonant points instead of computing the amplitudes.

\begin{figure}[hbpt]
	\centering
	\includegraphics[scale=0.3]{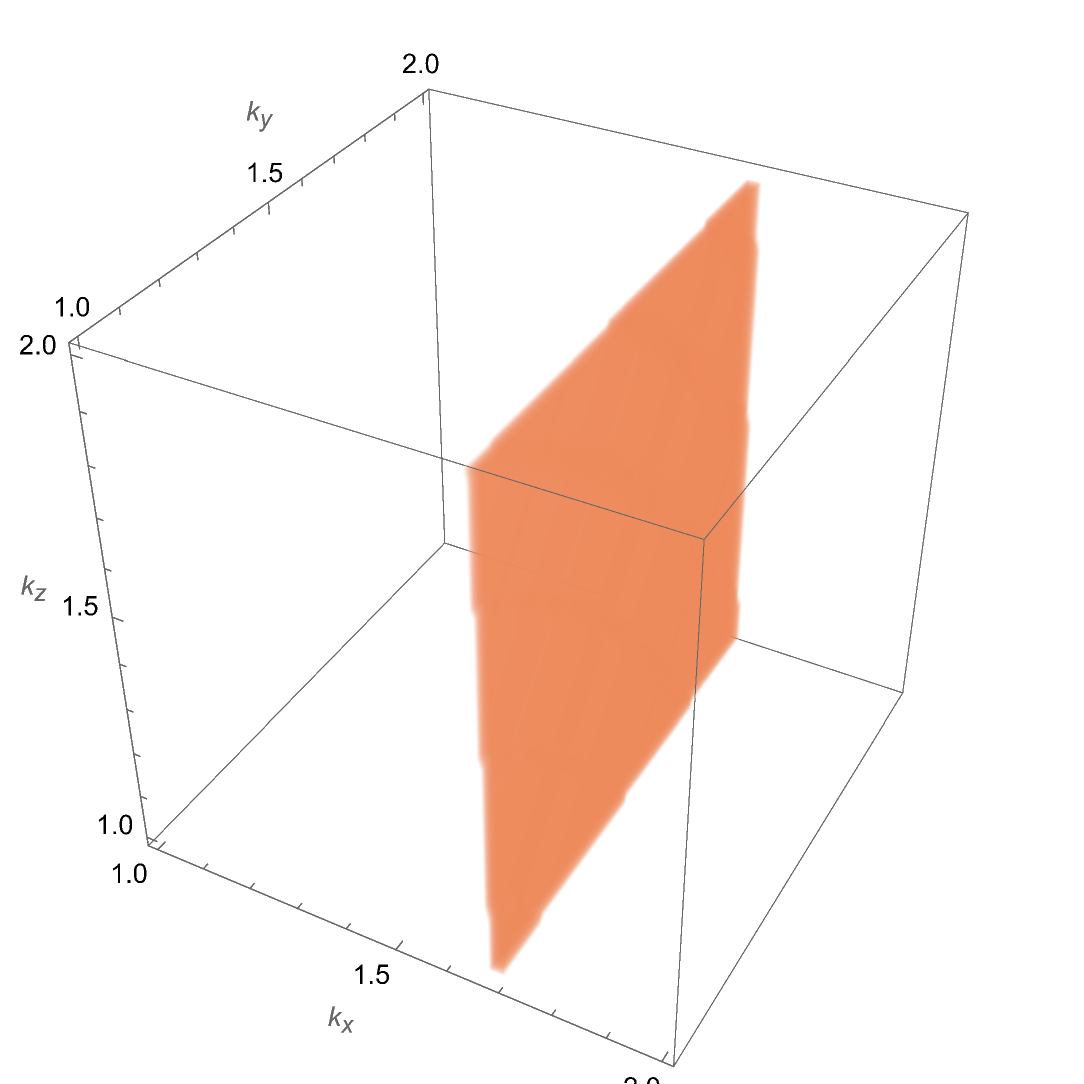}
	\caption{The two tensor modes $1$ and $1c$ resonate on a surface located approximately at $k_x=1.7$.
	}\label{TTVres}
\end{figure}

As in the AdS$_4$ case, we choose the wave vector of the quadratic mode to be along the $x$ direction and set $\vec{k}_1=(k_{x},k_{y},k_{z})$, $\vec{k}_2=(k_{x},-k_{y},-k_{z})$.
We check the resonance condition by computing
\begin{equation}
\text{Abs}\left(\omega^{(1)}_n(|\vec{k}_1|)+\omega^{(1)}_m(|\vec{k}_2|)-\omega^{(1)}_l(|\vec{k}_1+\vec{k}_2|)\right)
\end{equation}
with chosen overtones $n$ and $m$ for two tensor modes and all possible overtones $l$ for the tensor, vector, or scalar modes . 
The result is that there is no divergence in $\text{Tensor} \times \text{Tensor} \rightarrow \text{Tensor}$ and $\text{Tensor} \times \text{Tensor} \rightarrow \text{Scalar}$ in the region of small wave vectors ($0<k_x<2, 0<k_y<2, 0<k_z<2$).
However, there are resonant divergences in the process of $\text{Tensor} \times \text{Tensor} \rightarrow \text{Vector}$ if two lowest damping tensor modes $1$ and $1c$ are taken as the source. 
In fact, they resonate on a surface as shown in Figure \ref{TTVres}.
These divergences are analogues of the third form of divergences in the AdS$_4$ case where the divergence occurs on a curve.

\bibliography{main}

\providecommand{\href}[2]{#2}\begingroup\raggedright\begin{thebibliography}{10}

\bibitem{Price:1994pm}
R.~H. Price and J.~Pullin, \emph{{Colliding black holes: The Close limit}},
  \href{http://dx.doi.org/10.1103/PhysRevLett.72.3297}{\emph{Phys. Rev. Lett.}
  {\bf 72} (1994) 3297--3300}, [\href{http://arxiv.org/abs/gr-qc/9402039}{{\tt
  gr-qc/9402039}}].

\bibitem{Berti:2007fi}
E.~Berti, V.~Cardoso, J.~A. Gonzalez, U.~Sperhake, M.~Hannam, S.~Husa et~al.,
  \emph{{Inspiral, merger and ringdown of unequal mass black hole binaries: A
  Multipolar analysis}},
  \href{http://dx.doi.org/10.1103/PhysRevD.76.064034}{\emph{Phys. Rev. D} {\bf
  76} (2007) 064034}, [\href{http://arxiv.org/abs/gr-qc/0703053}{{\tt
  gr-qc/0703053}}].

\bibitem{Zerilli:1970wzz}
F.~J. Zerilli, \emph{{Gravitational field of a particle falling in a
  schwarzschild geometry analyzed in tensor harmonics}},
  \href{http://dx.doi.org/10.1103/PhysRevD.2.2141}{\emph{Phys. Rev. D} {\bf 2}
  (1970) 2141--2160}.

\bibitem{Davis:1971gg}
M.~Davis, R.~Ruffini, W.~H. Press and R.~H. Price, \emph{{Gravitational
  radiation from a particle falling radially into a schwarzschild black hole}},
  \href{http://dx.doi.org/10.1103/PhysRevLett.27.1466}{\emph{Phys. Rev. Lett.}
  {\bf 27} (1971) 1466--1469}.

\bibitem{Berti:2010ce}
E.~Berti, V.~Cardoso, T.~Hinderer, M.~Lemos, F.~Pretorius, U.~Sperhake et~al.,
  \emph{{Semianalytical estimates of scattering thresholds and gravitational
  radiation in ultrarelativistic black hole encounters}},
  \href{http://dx.doi.org/10.1103/PhysRevD.81.104048}{\emph{Phys. Rev. D} {\bf
  81} (2010) 104048}, [\href{http://arxiv.org/abs/1003.0812}{{\tt 1003.0812}}].

\bibitem{Vishveshwara:1970zz}
C.~V. Vishveshwara, \emph{{Scattering of Gravitational Radiation by a
  Schwarzschild Black-hole}},
  \href{http://dx.doi.org/10.1038/227936a0}{\emph{Nature} {\bf 227} (1970)
  936--938}.

\bibitem{Press:1971wr}
W.~H. Press, \emph{{Long Wave Trains of Gravitational Waves from a Vibrating
  Black Hole}}, \href{http://dx.doi.org/10.1086/180849}{\emph{Astrophys. J.
  Lett.} {\bf 170} (1971) L105--L108}.

\bibitem{Kokkotas:1999bd}
K.~D. Kokkotas and B.~G. Schmidt, \emph{{Quasinormal modes of stars and black
  holes}}, \href{http://dx.doi.org/10.12942/lrr-1999-2}{\emph{Living Rev. Rel.}
  {\bf 2} (1999) 2}, [\href{http://arxiv.org/abs/gr-qc/9909058}{{\tt
  gr-qc/9909058}}].

\bibitem{Nollert:1999ji}
H.-P. Nollert, \emph{{TOPICAL REVIEW: Quasinormal modes: the characteristic
  `sound' of black holes and neutron stars}},
  \href{http://dx.doi.org/10.1088/0264-9381/16/12/201}{\emph{Class. Quant.
  Grav.} {\bf 16} (1999) R159--R216}.

\bibitem{Berti:2009kk}
E.~Berti, V.~Cardoso and A.~O. Starinets, \emph{{Quasinormal modes of black
  holes and black branes}},
  \href{http://dx.doi.org/10.1088/0264-9381/26/16/163001}{\emph{Class. Quant.
  Grav.} {\bf 26} (2009) 163001}, [\href{http://arxiv.org/abs/0905.2975}{{\tt
  0905.2975}}].

\bibitem{Konoplya:2011qq}
R.~A. Konoplya and A.~Zhidenko, \emph{{Quasinormal modes of black holes: From
  astrophysics to string theory}},
  \href{http://dx.doi.org/10.1103/RevModPhys.83.793}{\emph{Rev. Mod. Phys.}
  {\bf 83} (2011) 793--836}, [\href{http://arxiv.org/abs/1102.4014}{{\tt
  1102.4014}}].

\bibitem{Cheung:2022rbm}
M.~H.-Y. Cheung et~al., \emph{{Nonlinear Effects in Black Hole Ringdown}},
  \href{http://dx.doi.org/10.1103/PhysRevLett.130.081401}{\emph{Phys. Rev.
  Lett.} {\bf 130} (2023) 081401}, [\href{http://arxiv.org/abs/2208.07374}{{\tt
  2208.07374}}].

\bibitem{Mitman:2022qdl}
K.~Mitman et~al., \emph{{Nonlinearities in Black Hole Ringdowns}},
  \href{http://dx.doi.org/10.1103/PhysRevLett.130.081402}{\emph{Phys. Rev.
  Lett.} {\bf 130} (2023) 081402}, [\href{http://arxiv.org/abs/2208.07380}{{\tt
  2208.07380}}].

\bibitem{Gleiser:1995gx}
R.~J. Gleiser, C.~O. Nicasio, R.~H. Price and J.~Pullin, \emph{{Second order
  perturbations of a Schwarzschild black hole}},
  \href{http://dx.doi.org/10.1088/0264-9381/13/10/001}{\emph{Class. Quant.
  Grav.} {\bf 13} (1996) L117--L124},
  [\href{http://arxiv.org/abs/gr-qc/9510049}{{\tt gr-qc/9510049}}].

\bibitem{Bruni:1996im}
M.~Bruni, S.~Matarrese, S.~Mollerach and S.~Sonego, \emph{{Perturbations of
  space-time: Gauge transformations and gauge invariance at second order and
  beyond}}, \href{http://dx.doi.org/10.1088/0264-9381/14/9/014}{\emph{Class.
  Quant. Grav.} {\bf 14} (1997) 2585--2606},
  [\href{http://arxiv.org/abs/gr-qc/9609040}{{\tt gr-qc/9609040}}].

\bibitem{Matarrese:1997ay}
S.~Matarrese, S.~Mollerach and M.~Bruni, \emph{{Second order perturbations of
  the Einstein-de Sitter universe}},
  \href{http://dx.doi.org/10.1103/PhysRevD.58.043504}{\emph{Phys. Rev. D} {\bf
  58} (1998) 043504}, [\href{http://arxiv.org/abs/astro-ph/9707278}{{\tt
  astro-ph/9707278}}].

\bibitem{Spiers:2023cip}
A.~Spiers, A.~Pound and J.~Moxon, \emph{{Second-order Teukolsky formalism in
  Kerr spacetime: Formulation and nonlinear source}},
  \href{http://dx.doi.org/10.1103/PhysRevD.108.064002}{\emph{Phys. Rev. D} {\bf
  108} (2023) 064002}, [\href{http://arxiv.org/abs/2305.19332}{{\tt
  2305.19332}}].

\bibitem{Spiers:2023mor}
A.~Spiers, A.~Pound and B.~Wardell, \emph{{Second-order perturbations of the
  Schwarzschild spacetime: Practical, covariant, and gauge-invariant
  formalisms}},
  \href{http://dx.doi.org/10.1103/PhysRevD.110.064030}{\emph{Phys. Rev. D} {\bf
  110} (2024) 064030}, [\href{http://arxiv.org/abs/2306.17847}{{\tt
  2306.17847}}].

\bibitem{Ioka:2007ak}
K.~Ioka and H.~Nakano, \emph{{Second and higher-order quasi-normal modes in
  binary black hole mergers}},
  \href{http://dx.doi.org/10.1103/PhysRevD.76.061503}{\emph{Phys. Rev. D} {\bf
  76} (2007) 061503}, [\href{http://arxiv.org/abs/0704.3467}{{\tt 0704.3467}}].

\bibitem{Nakano:2007cj}
H.~Nakano and K.~Ioka, \emph{{Second Order Quasi-Normal Mode of the
  Schwarzschild Black Hole}},
  \href{http://dx.doi.org/10.1103/PhysRevD.76.084007}{\emph{Phys. Rev. D} {\bf
  76} (2007) 084007}, [\href{http://arxiv.org/abs/0708.0450}{{\tt 0708.0450}}].

\bibitem{Lagos:2022otp}
M.~Lagos and L.~Hui, \emph{{Generation and propagation of nonlinear quasinormal
  modes of a Schwarzschild black hole}},
  \href{http://dx.doi.org/10.1103/PhysRevD.107.044040}{\emph{Phys. Rev. D} {\bf
  107} (2023) 044040}, [\href{http://arxiv.org/abs/2208.07379}{{\tt
  2208.07379}}].

\bibitem{Perrone:2023jzq}
D.~Perrone, T.~Barreira, A.~Kehagias and A.~Riotto, \emph{{Non-linear black
  hole ringdowns: An analytical approach}},
  \href{http://dx.doi.org/10.1016/j.nuclphysb.2023.116432}{\emph{Nucl. Phys. B}
  {\bf 999} (2024) 116432}, [\href{http://arxiv.org/abs/2308.15886}{{\tt
  2308.15886}}].

\bibitem{Bucciotti:2023ets}
B.~Bucciotti, A.~Kuntz, F.~Serra and E.~Trincherini, \emph{{Nonlinear
  quasi-normal modes: uniform approximation}},
  \href{http://dx.doi.org/10.1007/JHEP12(2023)048}{\emph{JHEP} {\bf 12} (2023)
  048}, [\href{http://arxiv.org/abs/2309.08501}{{\tt 2309.08501}}].

\bibitem{Khera:2023oyf}
N.~Khera, A.~Ribes~Metidieri, B.~Bonga, X.~Jim\'enez~Forteza, B.~Krishnan,
  E.~Poisson et~al., \emph{{Nonlinear Ringdown at the Black Hole Horizon}},
  \href{http://dx.doi.org/10.1103/PhysRevLett.131.231401}{\emph{Phys. Rev.
  Lett.} {\bf 131} (2023) 231401}, [\href{http://arxiv.org/abs/2306.11142}{{\tt
  2306.11142}}].

\bibitem{Bucciotti:2024zyp}
B.~Bucciotti, L.~Juliano, A.~Kuntz and E.~Trincherini, \emph{{Quadratic
  quasinormal modes of a Schwarzschild black hole}},
  \href{http://dx.doi.org/10.1103/PhysRevD.110.104048}{\emph{Phys. Rev. D} {\bf
  110} (2024) 104048}, [\href{http://arxiv.org/abs/2405.06012}{{\tt
  2405.06012}}].

\bibitem{Bourg:2024jme}
P.~Bourg, R.~Panosso~Macedo, A.~Spiers, B.~Leather, B.~Bonga and A.~Pound,
  \emph{{Quadratic quasi-normal mode dependence on linear mode parity}},
  \href{http://arxiv.org/abs/2405.10270}{{\tt 2405.10270}}.

\bibitem{Bucciotti:2024jrv}
B.~Bucciotti, L.~Juliano, A.~Kuntz and E.~Trincherini, \emph{{Amplitudes and
  polarizations of quadratic quasi-normal modes for a Schwarzschild black
  hole}}, \href{http://dx.doi.org/10.1007/JHEP09(2024)119}{\emph{JHEP} {\bf 09}
  (2024) 119}, [\href{http://arxiv.org/abs/2406.14611}{{\tt 2406.14611}}].

\bibitem{Kehagias:2024sgh}
A.~Kehagias and A.~Riotto, \emph{{Nonlinear Effects in Black Hole Ringdown Made
  Simple: Quasi-Normal Modes as Adiabatic Modes}},
  \href{http://arxiv.org/abs/2411.07980}{{\tt 2411.07980}}.

\bibitem{Ma:2024qcv}
S.~Ma and H.~Yang, \emph{{Excitation of quadratic quasinormal modes for Kerr
  black holes}},
  \href{http://dx.doi.org/10.1103/PhysRevD.109.104070}{\emph{Phys. Rev. D} {\bf
  109} (2024) 104070}, [\href{http://arxiv.org/abs/2401.15516}{{\tt
  2401.15516}}].

\bibitem{Khera:2024yrk}
N.~Khera, S.~Ma and H.~Yang, \emph{{Quadratic Mode Couplings in Rotating Black
  Holes and Their Detectability}},  \href{http://arxiv.org/abs/2410.14529}{{\tt
  2410.14529}}.

\bibitem{Maldacena:1997re}
J.~M. Maldacena, \emph{{The Large N limit of superconformal field theories and
  supergravity}},
  \href{http://dx.doi.org/10.4310/ATMP.1998.v2.n2.a1}{\emph{Adv. Theor. Math.
  Phys.} {\bf 2} (1998) 231--252},
  [\href{http://arxiv.org/abs/hep-th/9711200}{{\tt hep-th/9711200}}].

\bibitem{Horowitz:1999jd}
G.~T. Horowitz and V.~E. Hubeny, \emph{{Quasinormal modes of AdS black holes
  and the approach to thermal equilibrium}},
  \href{http://dx.doi.org/10.1103/PhysRevD.62.024027}{\emph{Phys. Rev. D} {\bf
  62} (2000) 024027}, [\href{http://arxiv.org/abs/hep-th/9909056}{{\tt
  hep-th/9909056}}].

\bibitem{Cardoso:2001bb}
V.~Cardoso and J.~P.~S. Lemos, \emph{{Quasinormal modes of Schwarzschild
  anti-de Sitter black holes: Electromagnetic and gravitational
  perturbations}},
  \href{http://dx.doi.org/10.1103/PhysRevD.64.084017}{\emph{Phys. Rev. D} {\bf
  64} (2001) 084017}, [\href{http://arxiv.org/abs/gr-qc/0105103}{{\tt
  gr-qc/0105103}}].

\bibitem{Cardoso:2001vs}
V.~Cardoso and J.~P.~S. Lemos, \emph{{Quasinormal modes of toroidal,
  cylindrical and planar black holes in anti-de Sitter space-times}},
  \href{http://dx.doi.org/10.1088/0264-9381/18/23/319}{\emph{Class. Quant.
  Grav.} {\bf 18} (2001) 5257--5267},
  [\href{http://arxiv.org/abs/gr-qc/0107098}{{\tt gr-qc/0107098}}].

\bibitem{Kovtun:2005ev}
P.~K. Kovtun and A.~O. Starinets, \emph{{Quasinormal modes and holography}},
  \href{http://dx.doi.org/10.1103/PhysRevD.72.086009}{\emph{Phys. Rev. D} {\bf
  72} (2005) 086009}, [\href{http://arxiv.org/abs/hep-th/0506184}{{\tt
  hep-th/0506184}}].

\bibitem{Miranda:2005qx}
A.~S. Miranda and V.~T. Zanchin, \emph{{Quasinormal modes of plane-symmetric
  anti-de Sitter black holes: A Complete analysis of the gravitational
  perturbations}},
  \href{http://dx.doi.org/10.1103/PhysRevD.73.064034}{\emph{Phys. Rev. D} {\bf
  73} (2006) 064034}, [\href{http://arxiv.org/abs/gr-qc/0510066}{{\tt
  gr-qc/0510066}}].

\bibitem{Miranda:2008vb}
A.~S. Miranda, J.~Morgan and V.~T. Zanchin, \emph{{Quasinormal modes of
  plane-symmetric black holes according to the AdS/CFT correspondence}},
  \href{http://dx.doi.org/10.1088/1126-6708/2008/11/030}{\emph{JHEP} {\bf 2008}
  (2008) 030}, [\href{http://arxiv.org/abs/0809.0297}{{\tt 0809.0297}}].

\bibitem{Morgan:2009pn}
J.~Morgan, V.~Cardoso, A.~S. Miranda, C.~Molina and V.~T. Zanchin,
  \emph{{Gravitational quasinormal modes of AdS black branes in d spacetime
  dimensions}},
  \href{http://dx.doi.org/10.1088/1126-6708/2009/09/117}{\emph{JHEP} {\bf 2009}
  (2009) 117}, [\href{http://arxiv.org/abs/0907.5011}{{\tt 0907.5011}}].

\bibitem{Birmingham:2001pj}
D.~Birmingham, I.~Sachs and S.~N. Solodukhin, \emph{{Conformal field theory
  interpretation of black hole quasinormal modes}},
  \href{http://dx.doi.org/10.1103/PhysRevLett.88.151301}{\emph{Phys. Rev.
  Lett.} {\bf 88} (2002) 151301},
  [\href{http://arxiv.org/abs/hep-th/0112055}{{\tt hep-th/0112055}}].

\bibitem{Son:2002sd}
D.~T. Son and A.~O. Starinets, \emph{{Minkowski space correlators in AdS / CFT
  correspondence: Recipe and applications}},
  \href{http://dx.doi.org/10.1088/1126-6708/2002/09/042}{\emph{JHEP} {\bf 09}
  (2002) 042}, [\href{http://arxiv.org/abs/hep-th/0205051}{{\tt
  hep-th/0205051}}].

\bibitem{Pantelidou:2022ftm}
C.~Pantelidou and B.~Withers, \emph{{Thermal three-point functions from
  holographic Schwinger-Keldysh contours}},
  \href{http://dx.doi.org/10.1007/JHEP04(2023)050}{\emph{JHEP} {\bf 04} (2023)
  050}, [\href{http://arxiv.org/abs/2211.09140}{{\tt 2211.09140}}].

\bibitem{Moncrief:1974am}
V.~Moncrief, \emph{{Gravitational perturbations of spherically symmetric
  systems. I. The exterior problem.}},
  \href{http://dx.doi.org/10.1016/0003-4916(74)90173-0}{\emph{Annals Phys.}
  {\bf 88} (1974) 323--342}.

\bibitem{Garat:1999vr}
A.~Garat and R.~H. Price, \emph{{Gauge invariant formalism for second order
  perturbations of Schwarzschild space-times}},
  \href{http://dx.doi.org/10.1103/PhysRevD.61.044006}{\emph{Phys. Rev. D} {\bf
  61} (2000) 044006}, [\href{http://arxiv.org/abs/gr-qc/9909005}{{\tt
  gr-qc/9909005}}].

\bibitem{Brizuela:2007zza}
D.~Brizuela, J.~M. Martin-Garcia and G.~A.~M. Marugan, \emph{{High-order
  gauge-invariant perturbations of a spherical spacetime}},
  \href{http://dx.doi.org/10.1103/PhysRevD.76.024004}{\emph{Phys. Rev. D} {\bf
  76} (2007) 024004}, [\href{http://arxiv.org/abs/gr-qc/0703069}{{\tt
  gr-qc/0703069}}].

\bibitem{Brizuela:2009qd}
D.~Brizuela, J.~M. Martin-Garcia and M.~Tiglio, \emph{{A Complete
  gauge-invariant formalism for arbitrary second-order perturbations of a
  Schwarzschild black hole}},
  \href{http://dx.doi.org/10.1103/PhysRevD.80.024021}{\emph{Phys. Rev. D} {\bf
  80} (2009) 024021}, [\href{http://arxiv.org/abs/0903.1134}{{\tt 0903.1134}}].

\bibitem{Regge:1957td}
T.~Regge and J.~A. Wheeler, \emph{{Stability of a Schwarzschild singularity}},
  \href{http://dx.doi.org/10.1103/PhysRev.108.1063}{\emph{Phys. Rev.} {\bf 108}
  (1957) 1063--1069}.

\bibitem{Zerilli:1970se}
F.~J. Zerilli, \emph{{Effective potential for even parity Regge-Wheeler
  gravitational perturbation equations}},
  \href{http://dx.doi.org/10.1103/PhysRevLett.24.737}{\emph{Phys. Rev. Lett.}
  {\bf 24} (1970) 737--738}.

\bibitem{Chou:1984es}
K.-c. Chou, Z.-b. Su, B.-l. Hao and L.~Yu, \emph{{Equilibrium and
  Nonequilibrium Formalisms Made Unified}},
  \href{http://dx.doi.org/10.1016/0370-1573(85)90136-X}{\emph{Phys. Rept.} {\bf
  118} (1985) 1--131}.

\bibitem{Meltzer:2021bmb}
D.~Meltzer, \emph{{Dispersion Formulas in QFTs, CFTs, and Holography}},
  \href{http://dx.doi.org/10.1007/JHEP05(2021)098}{\emph{JHEP} {\bf 05} (2021)
  098}, [\href{http://arxiv.org/abs/2103.15839}{{\tt 2103.15839}}].

\bibitem{Amado:2007pv}
I.~Amado, C.~Hoyos-Badajoz, K.~Landsteiner and S.~Montero, \emph{{Absorption
  lengths in the holographic plasma}},
  \href{http://dx.doi.org/10.1088/1126-6708/2007/09/057}{\emph{JHEP} {\bf 09}
  (2007) 057}, [\href{http://arxiv.org/abs/0706.2750}{{\tt 0706.2750}}].

\bibitem{Bizon:2011gg}
P.~Bizon and A.~Rostworowski, \emph{{On weakly turbulent instability of anti-de
  Sitter space}},
  \href{http://dx.doi.org/10.1103/PhysRevLett.107.031102}{\emph{Phys. Rev.
  Lett.} {\bf 107} (2011) 031102}, [\href{http://arxiv.org/abs/1104.3702}{{\tt
  1104.3702}}].

\bibitem{Jalmuzna:2011qw}
J.~Jalmuzna, A.~Rostworowski and P.~Bizon, \emph{{A Comment on AdS collapse of
  a scalar field in higher dimensions}},
  \href{http://dx.doi.org/10.1103/PhysRevD.84.085021}{\emph{Phys. Rev. D} {\bf
  84} (2011) 085021}, [\href{http://arxiv.org/abs/1108.4539}{{\tt 1108.4539}}].

\bibitem{Gubser:2008px}
S.~S. Gubser, \emph{{Breaking an Abelian gauge symmetry near a black hole
  horizon}}, \href{http://dx.doi.org/10.1103/PhysRevD.78.065034}{\emph{Phys.
  Rev. D} {\bf 78} (2008) 065034}, [\href{http://arxiv.org/abs/0801.2977}{{\tt
  0801.2977}}].

\bibitem{Hartnoll:2008vx}
S.~A. Hartnoll, C.~P. Herzog and G.~T. Horowitz, \emph{{Building a Holographic
  Superconductor}},
  \href{http://dx.doi.org/10.1103/PhysRevLett.101.031601}{\emph{Phys. Rev.
  Lett.} {\bf 101} (2008) 031601}, [\href{http://arxiv.org/abs/0803.3295}{{\tt
  0803.3295}}].

\bibitem{Hartnoll:2009ns}
S.~A. Hartnoll, J.~Polchinski, E.~Silverstein and D.~Tong, \emph{{Towards
  strange metallic holography}},
  \href{http://dx.doi.org/10.1007/JHEP04(2010)120}{\emph{JHEP} {\bf 04} (2010)
  120}, [\href{http://arxiv.org/abs/0912.1061}{{\tt 0912.1061}}].

\bibitem{Grandclement:2007sb}
P.~Grandclement and J.~Novak, \emph{{Spectral methods for numerical
  relativity}}, \href{http://dx.doi.org/10.12942/lrr-2009-1}{\emph{Living Rev.
  Rel.} {\bf 12} (2009) 1}, [\href{http://arxiv.org/abs/0706.2286}{{\tt
  0706.2286}}].

\bibitem{Monteiro:2009ke}
R.~Monteiro, M.~J. Perry and J.~E. Santos, \emph{{Semiclassical instabilities
  of Kerr-AdS black holes}},
  \href{http://dx.doi.org/10.1103/PhysRevD.81.024001}{\emph{Phys. Rev. D} {\bf
  81} (2010) 024001}, [\href{http://arxiv.org/abs/0905.2334}{{\tt 0905.2334}}].

\bibitem{Jansen:2017oag}
A.~Jansen, \emph{{Overdamped modes in Schwarzschild-de Sitter and a Mathematica
  package for the numerical computation of quasinormal modes}},
  \href{http://dx.doi.org/10.1140/epjp/i2017-11825-9}{\emph{Eur. Phys. J. Plus}
  {\bf 132} (2017) 546}, [\href{http://arxiv.org/abs/1709.09178}{{\tt
  1709.09178}}].

\end{thebibliography}\endgroup
\bibliographystyle{JHEP}

\end{document}